\documentclass[preprint]{aastex}

\shorttitle{RSGs in the LMC: I. The P-L Relation} \shortauthors{Yang \& Jiang}

\begin{document}

\title{Red Supergiant Stars in the Large Magellanic Cloud:\\
 I. The Period-Luminosity Relation}

\author{Ming Yang and B.~W. Jiang}
\affil{Department of Astronomy, Beijing Normal University, Beijing 100875, China; {myang@mail.bnu.edu.cn,
bjiang@bnu.edu.cn}}

\begin{abstract}

From previous samples of Red Supergiants (RSGs) by various groups,
191 objects are assembled to compose a large sample of RSG
candidates in LMC. For 189 of them, the identity as a RSG is
verified by their brightness and color indexes in several near- and
mid-infrared bands related to the 2MASS JHKs bands and the
\emph{Spitzer}/IRAC and \emph{Spitzer}/MIPS bands. From the visual
time-series photometric observations by the ASAS and MACHO projects
which cover nearly 8-10 years, the period and amplitude of light
variation are analyzed carefully using both the PDM and Period04
methods. According to the properties of light variation, these
objects are classified into five categories: (1) 20 objects are
saturated in photometry or located in crowded stellar field with
poor photometric results, (2) 35 objects with too complex variation
to have any certain period, (3) 23 objects with irregular variation,
(4) 16 objects with semi-regular variation, and (5) 95 objects with
Long Secondary Period (LSP) among which 31 have distinguishable
short period, and 51 have a long period shorter than 3000 days
that can be determined with reasonable accuracy.
For the semi-regular variables and the LSP variables with
distinguishable short period, the period-luminosity relation is
analyzed in the visual, near-infrared and mid-infrared bands. It is
found that the P-L relation is tight in the infrared bands such as
the 2MASS JHKs bands and the \emph{Spitzer}/IRAC bands, in
particular in the \emph{Spitzer}/IRAC [3.6] and [4.5] bands;
meanwhile, the P-L relation is relatively sparse in the V band which
may be caused by the inhomogeneous interstellar extinction.
The results are compared with others' P-L relationships for RSGs and
the P-L sequences of red giants in LMC.

\end{abstract}


\section{Introduction}

Red SuperGiants (RSGs) are evolved, He-burning, extreme Population I
stars with moderately high mass (10-25\,$M_{\sun}$) and a degenerate
core. They have very large radii (200-1500\,$R_{\sun}$)
\citep{Levesque05}, and spectral types of M or late-K ($T_{\rm
eff}\approx 3000\sim4000{\rm K}$) \citep{Massey08}. The very large
radii makes them one class of the most luminous stars. Besides, they
have large mass loss rate (MLR) which produces strong stellar wind
or even superwind from the outer layer and creates a dusty envelope
around the central star. For this reason, RSGs contribute to the
interstellar medium, nucleosynthesis and chemical evolution of
galaxies \citep{Fusi76, Reimers77, Chiosi78, Stothers79, Maeder81,
Chiosi86, MacGregor92}.

One interesting characteristic of RSGs is that they show optical
variability with relatively long period. Some of them are
semi-regular variable and referred to as Long Period Variable stars
(LPVs). Indeed, LPVs generally include two major types: the
Asymptotic Giant Branch stars (AGBs) and the RSGs. Both are luminous
with $M_{\rm bol}\leqslant-6$ \citep{Wood83} and have a variation
period ranging from several hundreds to thousand days. The AGB stars
often have  large amplitude and well-pronounced periodicity while
RSGs have small amplitude and not so regular periodicity. A lot of
research has been devoted to the study of the long-term variability
of RSGs and divided them ambiguously into two groups, the
semi-regular and the irregular. The division is ambiguous mostly
because of the blurred boundary between the semi-regular and the
irregular behavior so that sometimes arbitrary decisions are made
for the boundary objects. In the General Catalogue of Variable Stars
(GCVS; \citet{Kholopov85}), SRc and Lc are the two types of RSGs
corresponding to the semi-regular and the irregular respectively.
Even SRc exhibits two kinds of variation. One is the
shorter-period variation characterized by a time scale of
several hundreds or less days and identified as the radial pulsation
at the fundamental, first and possibly second overtone mode
 \citep{Stothers69, Wood83, Lovy84, Schaller90, Li94, Heger97,
Guo02}. The other is the longer-period variation with a
period longer than 1000 days, also known as Long Secondary Period
(LSP) and found additionally in AGBs and Red Giant stars (RGs)
\citep{Stothers72, Percy99, Olivier03, Derekas06, Soszynski07,
Fraser08, Wood09}. The mechanism for this LSP is still
unknown. The models like binary, pulsation, convection cell and
surface hot spot are proposed but none of them agrees with all the
observations and theoretical expectations \citep{Buscher90,
Wilson92, Tuthill97, Groenewegen04, Wood04, Kiss06, Messina07,
Haubois09, Nie10}. Concerning Lc, the irregular variation is
considered to be caused by large convection cells which can account
for the entire or part of the variation. This scenario is consistent
with the profile of the light curve that has an irregular pattern
overlying a regular pattern \citep{Schwarzschild75, Antia84,
Kiss06}.

The difficulty of observing RSGs is the very long time scale of
variation. Without multi-cycle coverage of light variation, it is
hard even to fully characterize their variation in luminosity, in
particular for those with the variation time scale as long as a few
thousands of days. Fortunately, with the help of robotic telescopes,
long-term photometric monitoring of RSGs is carried out in the past
decade which gives us an opportunity to investigate their light
variation with a more solid observational base. \citet{Kiss06} have
done a comprehensive study of 48 Galactic RSGs using long-term
visual light curves collected by the American Association of
Variable Star Observers (AAVSO) with a full span of time of about 60
years. They found semi-regular, irregular variables and LSPs in
their targets. Very recently, using a 10-year photometric
monitoring dataset by All Sky Automated Survey (ASAS),
\citet{Szczygiel10} reported their work of discovering 85 semi- or
non-periodic RSGs in LMC which are included in our study.

The period-luminosity (P-L) relation of RSGs has aroused much
interest. The great intrinsic luminosity of RSGs makes them bright
enough to be observed even in distant galaxies. \citet{Glass79}
first discussed the potential of using RSGs as an extragalactic
distance indicator. Later, infrared surveys of RSGs in the LMC and
SMC by \citet{Feast80} and \citet{Catchpole81} yielded a rough P-L
relation. \citet{Wood83} also made use of infrared JHK photometry
and low-dispersion red spectrum to investigate the LPVs in the LMC
and SMC. They suggested a criterion to distinguish AGB
stars and RSGs by $M_{\rm bol}=-7.1$, which needs
reexamination since some super-AGB stars can be so bright
as $M_{\rm bol}$ is up to -8\,mag \citep{Poelarends08}.
Nevertheless, they found that RSGs follow a P-L sequence which is
approximately one magnitude brighter than AGBs in the \emph{K} band.
Several other groups have studied the P-L relation of RSGs in
multiple bands at different distance scales such as in the Milky
Way, LMC, SMC and M33 \citep{Feast80, Catchpole81, Kinman87,
Pierce00, Kiss06}. \citet{Pierce00} re-calibrated the RSG P-L
relations in Per OB1, LMC and M33 in various bands and suggested a
uniform relation in these heterogeneous environments which can be
used to measure the distance to M101. Meanwhile, \citet{Kiss06},
based on the analysis of 48 Galactic RSGs with almost 60-years'
data, obtained a P-L relation which is similar to that of AGB stars
in LMC.  The conclusions from various researchers have some
discrepancies, which could be caused by the volume of the sample,
the accuracy of the periods or the indicator of the luminosity. In
present work, we re-analyze the P-L relation of RSGs in LMC, by
carefully selecting a large sample, and by using the most up-to-date
photometric data and a few new luminosity indicators.

\section{Sample Selection and Data Analysis}

First of all, to have a pure and as large as possible sample is the
basis to determine a reliable P-L relationship. But this is a
difficult task, because RSGs are easily confused with AGB stars
since both are red and luminous. Previous works selected the sample
of RSGs in LMC by non-uniform criteria. Early in 1980,
\citet{Feast80} identified 24 RSGs from their period and luminosity
based on the old Harvard work which included 7 stars from
\cite{Glass79}. \citet{Wood83} produced a new catalog of 26 sources
by a detailed analysis of their near-infrared photometry and
low-dispersion red spectrum. This catalog became the resource of
\citet{Pierce00} to make their list of 24 RSGs after excluding those
which could be AGB stars with a period shorter than 400 days. The
sample was largely extended by \citet{Massey03} to be consisted of
158 sources through multi-object spectroscopy of a sample of red
stars identified by \cite{Massey02}. They simultaneously made use of
high-accuracy (${\rm <1\,km\,s^{-1}}$) radial velocities for all the
candidates to confirm them as RSGs. Recently, \citet{Kastner08}
selected another sample in a new way, i.e. by choosing the most
mid-IR-luminous stars from the 2MASS-\emph{MSX}-\emph{Spitzer}
photometric surveys, meanwhile \citet{Buchanan09} had a supplement
to this paper that identified additional seven objects as RSGs via
their \emph{Spitzer} spectral features and luminosity.

To make a sample as complete as possible, we compiled the samples
from \citet{Feast80}, \citet{Pierce00}, \citet{Massey03}, and
\citet{Kastner08} altogether as a first step. This preliminary
catalog contains in total 232 objects.  Actually, these
samples overlap, so we adopt the sources by the order of
publication to subtract previous sources from following papers.
After this,  the sample is consisted of 200 stars, specifically 23
from \citet{Feast80}, 11 from \citet{Pierce00}, 140 from
\citet{Massey03}, and 26 from \citet{Kastner08}.

For the light variation, the time-series photometric data in the
visual bands are taken from the databases of the All Sky Automated
Survey (ASAS) \citep{Pojmanski02} and the MAssive Compact Halo
Objects (MACHO) projects \citep{Alcock97}. Although the Optical
Gravitational Lensing Experiment (OGLE) \citep{Szymanski05} seems to
be a good resource as it also observes  LMC for several years, it
does not provide useful photometric data due to that all targets are
saturated in its I band. The one-epoch near- and mid-infrared
photometric data are retrieved from the Two Micron All Sky
Survey(2MASS) \citep{Skrutskie06} PSC, and the \textit{Spitzer}/SAGE
Legacy Program \citep{Meixner06} database.

\subsection{Color-Magnitude Diagrams and Two-Color Diagrams}

Because RSGs are easily confused with bright AGB stars and blue
supergiants if judged only from the brightness, we need to
examine the sample to make sure that every source in the
sample is a true RSG. To identify the RSGs, the basic criterion is
the brightness, and modified by the effective temperature. Thus the
color-magnitude diagrams (CMDs) and two-color diagrams (TCDs) are
the tools. Although they should be red and luminous, RSGs always
have heavy dusty envelopes due to large mass loss rate which cause
large extinction at shorter wavelengths but much less at longer
wavelengths. To avoid the effect of extinction as much as possible,
we choose the near- and mid-infrared bands of 2MASS and
\emph{Spitzer}/SAGE where the extinction is much smaller than in the
optical bands. For the infrared magnitudes, the SAGE Winter '08 IRAC
Catalog is selected because it includes the Epoch 1 and Epoch 2 (the
SAGE project is composed of two-epoch observations) IRAC images and
is already cross-associated with the 2MASS Point Source Catalog
\citep{Cutri04}. This is a highly reliable catalog as a subset of
the IRAC Archive catalog through strict selection. In addition, we
also use the SAGE Winter '08 MIPS 24$\mu$m Catalog that is
cross-associated with the SAGE Winter '08 IRAC catalog. More details
about the SAGE catalogs can be found in \citet{Meixner06} and the
SAGE Data Description
Document\footnote{http://irsa.ipac.caltech.edu/data/SPITZER/SAGE/doc/}.

\begin{figure}
\centering
\includegraphics[width=\textwidth, bb=20 20 575 575]{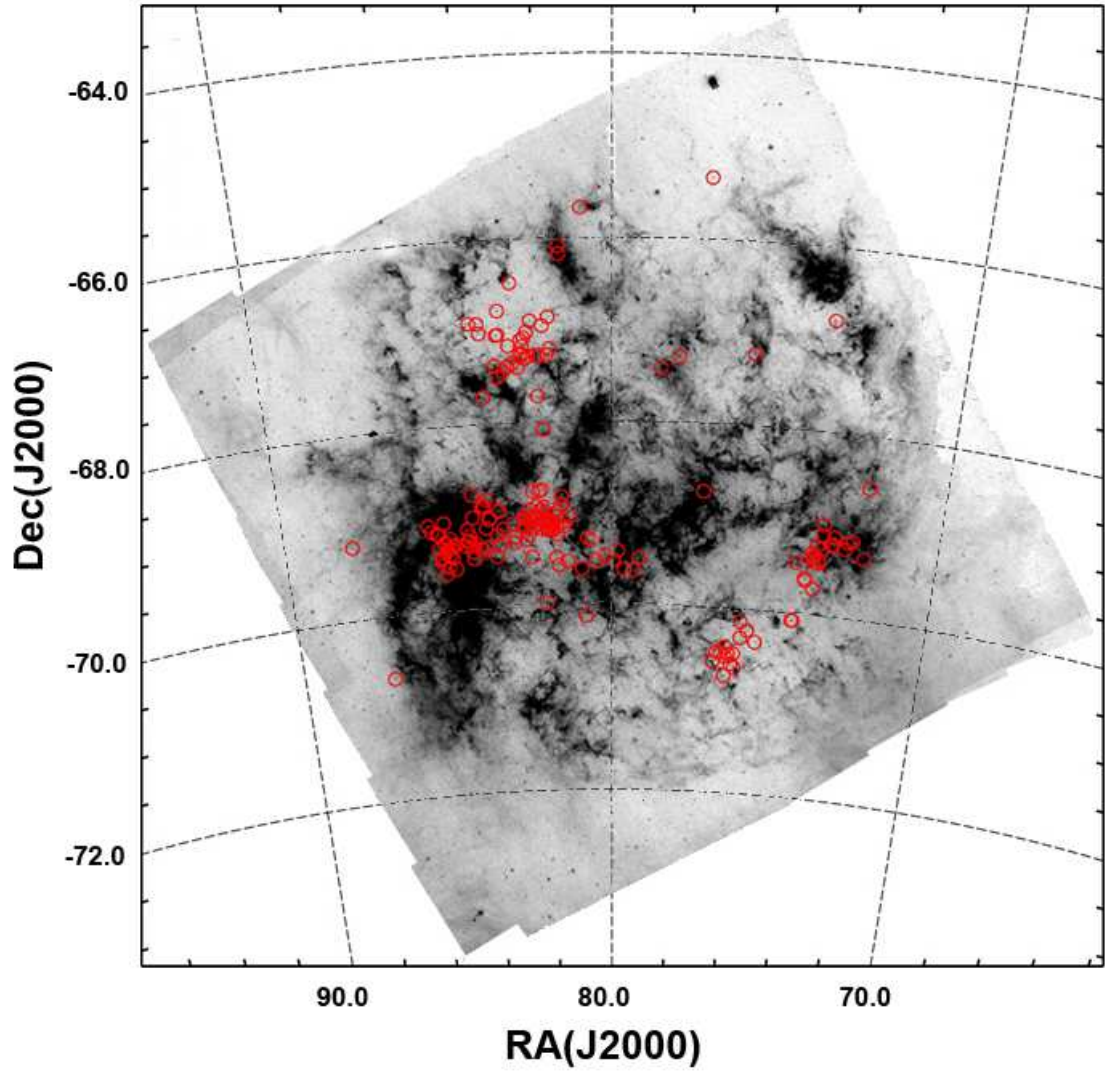}
\caption{
    The spatial distribution of all the 191 sample stars superposed on the \emph{Spitzer}/SAGE 8$\mu$m
    mosaic image. Many stars clump near the 30 Doradus area.
} \label{lmc}
\end{figure}

The RSG candidates are cross-identified in the SAGE catalog by a
1$\arcsec$  search radius that coincides with the nominal pointing
accuracy of \emph{Spitzer} and choosing the closest and brightest
counterpart from the objects within the search circle. Actually,
because RSGs are of great intrinsic luminosity and located always in
sparse stellar field, the 1$\arcsec$ search radius only results in
one counterpart for each source. In addition, even the source  has
null bands in the 2MASS/JHKs or \emph{Spitzer}/IRAC bands, it is
retained other than dropped.  Nine targets which have no counterpart
in the infrared catalog within the search circle are dropped. As the
final SAGE catalogue is yet to be released, the Epoch 1 and Epoch 2
data are still separated, but the differences in the infrared
magnitudes between them are less than one percent and negligible.
Therefore, we used the Epoch which has more sources and extracted
the rest from the other Epoch. For example, the Massey sample has
more sources in Epoch 2 in the IRAC bands, while more sources in
Epoch 1 in the MIPS bands. As a result, all the sources have
photometric measurements for the integral wavelength range covering from
the J (1.2$\mu$m) band to the MIPS 24$\mu$m band that would give a
better identification and comprehensive view of infrared properties
of RSGs. Finally, the sample is consisted of 191 sources.
Fig.~\ref{lmc} shows the spatial distribution of all the 191 stars
superposed on the \emph{Spitzer}/SAGE 8 $\mu$m mosaic image. We note
that many stars clump near the 30 Doradus area. Table \ref{191tab}
lists their coordinates and infrared magnitudes, where the '---'
symbol denotes the data is missing in the corresponding band, in
total, 28 measurements (with 13 at 24$\mu$m) are missing for 21
objects, as well as the resources for photometric data and the
reference.

For comparison, we add the 1268 massive stars (M $\geqslant$
8$M_{\sun}$) in the LMC from \citet{Bonanos09}. They are
collected from literatures and have been identified by the same
criteria in the SAGE infrared data as ours.

\begin{figure}
\centering
\includegraphics[width=\textwidth, bb=50 280 550 750]{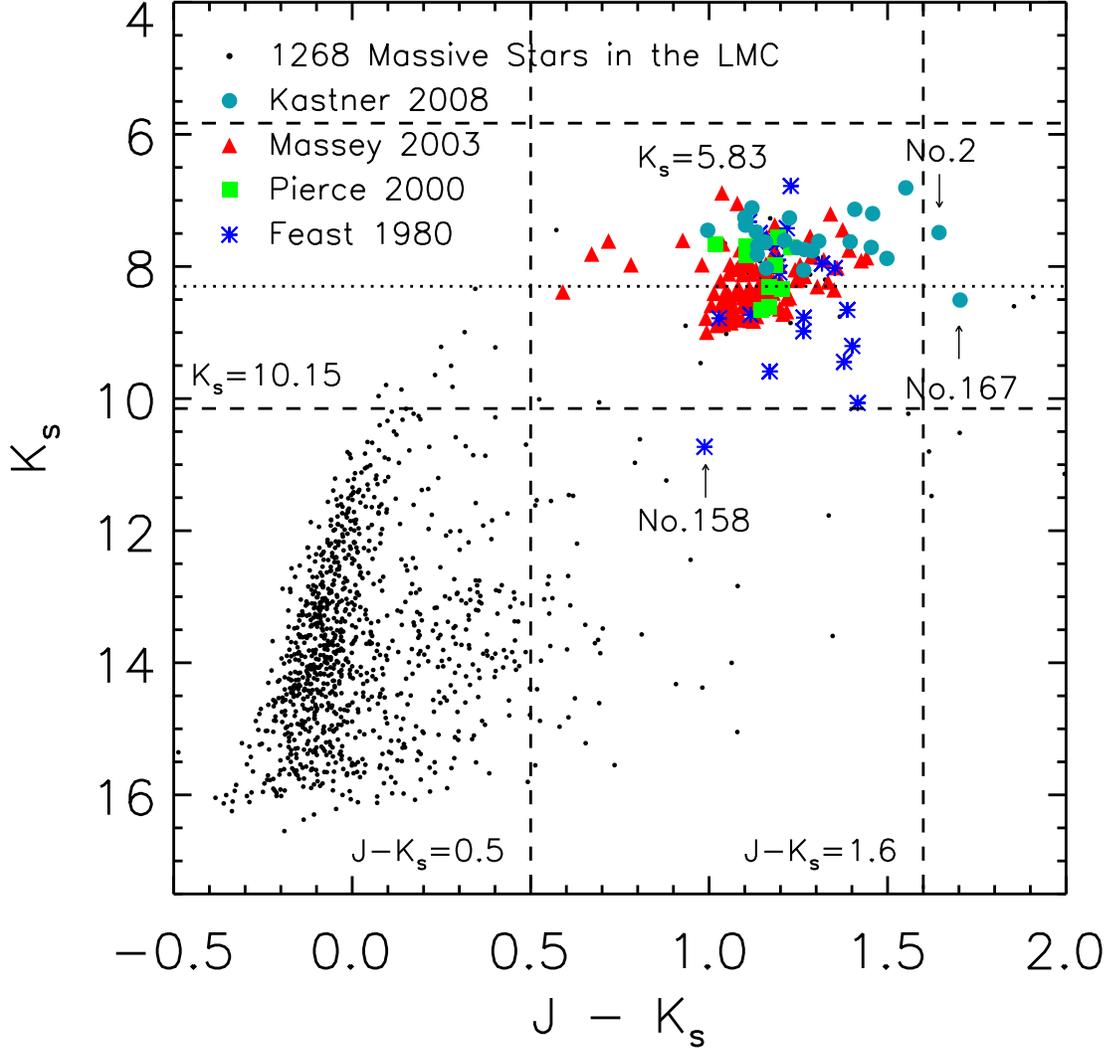}
\caption{
    $[K_{\rm S}]$ vs. $J - [K_{\rm S}]$ CMD for all targets. For comparison,
    the 1268 massive
    stars from \citet{Bonanos09} are added as background and denoted by black dots.
    Different symbols represent different
    resources. Most of the targets clump at $J - K_{\rm S} = 0.5 \sim 1.6$. The
    vertical
    dashed line with $J - K_{\rm S} = 1.6$ gives the boundary of
    carbon-rich star \citep{Hughes90} and
    another with $J - K_{\rm S} = 0.5$  the observational boundary
    of RSGs in LMC \citep{Josselin00}.
    The horizontal dashed lines with $K_{\rm S}=5.83$ and $K_{\rm S}=10.15$
    show the theoretical
    luminosity
    boundaries in the $K_{\rm S}$ band (see the text for details). The dotted line
    shows the criterion of $M_{\rm bol}=-7.1$ to
    distinguish the
    AGB and RSG stars defined by \citet{Wood83}.
} \label{fig1}
\end{figure}

Fig.~\ref{fig1} is the $J-K_{\rm S}$/$K_{\rm S}$ CMD for all
the targets. No interstellar reddening is taken into account, as the
extinction in the J and $K_{\rm S}$ bands is only about 0.2 and
0.06 mag respectively if the suggested E(B-V)=0.2 is adopted,
comparable to the observational uncertainty. No.53 and No.178
(the ID number in Table~\ref{191tab}) without the J and
$K_{\rm S}$ band data are absent in this diagram. In
Fig.~\ref{fig1}, most RSG candidates locate within a region of
6.5 $< K_{\rm S} <$ 10 and 0.5 $< J-K_{\rm S} < $ 1.6 which
corresponds to the H region of \citet{Nikolaev00} for the LMC K$\sim$M supergiants but has a higher tip. This means that our
sample extends to more luminous sources.

For identification, we set the boundaries of luminosity and color
index. According to the mass-luminosity relation for massive stars
($9-30M_{\sun}$), $L/L_{\sun}=(M/M_{\sun})^\gamma$ with $\gamma$
very close to 4.0 \citep{Stothers71}, the luminosity range of RSGs
is $10^4-25^4 L_{\sun}$ with the mass range $10-25 M_{\sun}$, which
can be converted to the absolute bolometric magnitude $M_{\rm
bol}=4.74-2.5\log(L/L_{\sun})$ to range from -9.58 to -5.26.
Furthermore, $M_{\rm bol}$ is converted to the K band magnitude from
$m_{bol}=m_{\rm K}+3$ \citep{Josselin00}. A distance modulus
of 18.41 mag \citep{Macri06} is used to convert the observed $K_{\rm
S}$ to the absolute magnitude. Because the difference of magnitude
between the K and $K_{\rm S}$ bands is very small and can be
ignored, we finally get the $K_{\rm S}$ band magnitude should be
between 5.83 mag and 10.15 mag, shown as the dashed horizontal lines
in Fig~\ref{fig1}. For the color index $J-K_{\rm S}$, the lower
limit is 0.5, the same as the lower observational boundary of
\citet{Josselin00}
 and the upper limit is 1.6, the same as the boundary of carbon-rich
stars defined by \citet{Hughes90}. These limits also are
consistent with the color indexes of most candidates.
Additionally marked by dotted line in Fig.~\ref{fig1} is the
criterion of $M_{\rm bol}=-7.1$ which was proposed to
distinguish the AGBs and RSGs by \citet{Wood83}. It can be seen
that this criterion would leave a third of our targets out and
difficult to reconcile with others. Besides clustering in
Fig.~\ref{fig1}, the RSGs exhibit an upward tendency toward the
red end in $J-K_{\rm S}$, which indicates a higher mass loss rate at
higher $K_{\rm S}$ luminosity. This tendency becomes clearer at
longer wavelengths in subsequent analysis. The broadening of
the sequence may be caused by different MLR. In this CMD, there
are a few candidates lying outside the boundaries. In the far
upright of this figure are No.2 and No.167, they exceed a
little bit the upper limit of $J-K_{\rm S}$ = 1.6. No.58 is too
faint to satisfy the lower limit of $K_{\rm S}$ band
luminosity.

From Fig.~\ref{fig1}, the identification of the RSG candidates
is almost finished. Meanwhile, RSGs are in the evolved phase
and have large MLR. Several authors observed that RSGs
have circumstellar dust features from 8$\mu$m to 12$\mu$m
\citep{Hagen78, Skinner88, Josselin00}, mainly the 9.7$\mu$m
silicate feature and the 12.1$\mu$m feature. Therefore, in
order to further confirm the identification of RSGs, the color
index involving a MIR band should be helpful.

\begin{figure}
\centering
\includegraphics[width=\textwidth, bb=50 280 550 750]{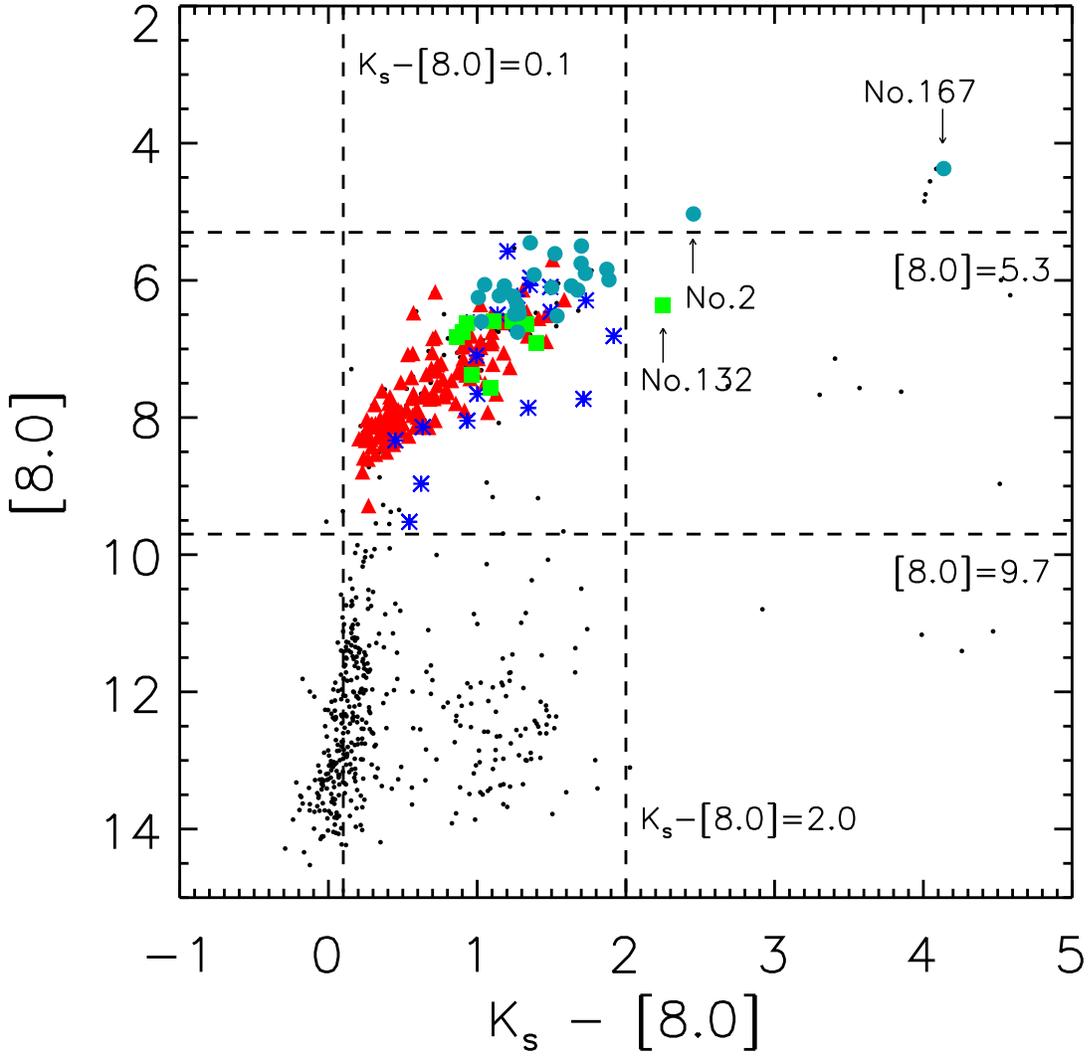}
\caption{
    The same as Fig.~\ref{fig1}, but for [8.0] vs. $K_{\rm S}$ - [8.0]. We set our own limits for RSGs to
    include
    98\% clumped targets: $5.3 \leq [8.0] \leq 9.7$ and $0.1 \leq K_{\rm S} - [8.0] \leq 2.0$.
} \label{fig2}
\end{figure}

In Fig.~\ref{fig2}, the targets are plotted in the $K_{\rm
S}$ - [8.0]/[8.0] CMD. Nos. 53, 58 and 178 lack  the $K_{\rm
S}$ and/or [8.0] band data, they are not present in this
diagram. Because no ready-made criterion can be used on these
bands, we set our own limits for RSGs to include 98\% clumped
targets, which means:  $5.3 \leq [8.0] \leq 9.7$ and $0.1 \leq
K_{\rm S} - [8.0] \leq 2.0$. Here are again a couple of outliers,
No.2 and No.167 are still in the far right and a new outlier
obviously redder than the others in the right side is No.132.

\begin{figure}
\centering
\includegraphics[width=\textwidth, bb=50 280 550 750]{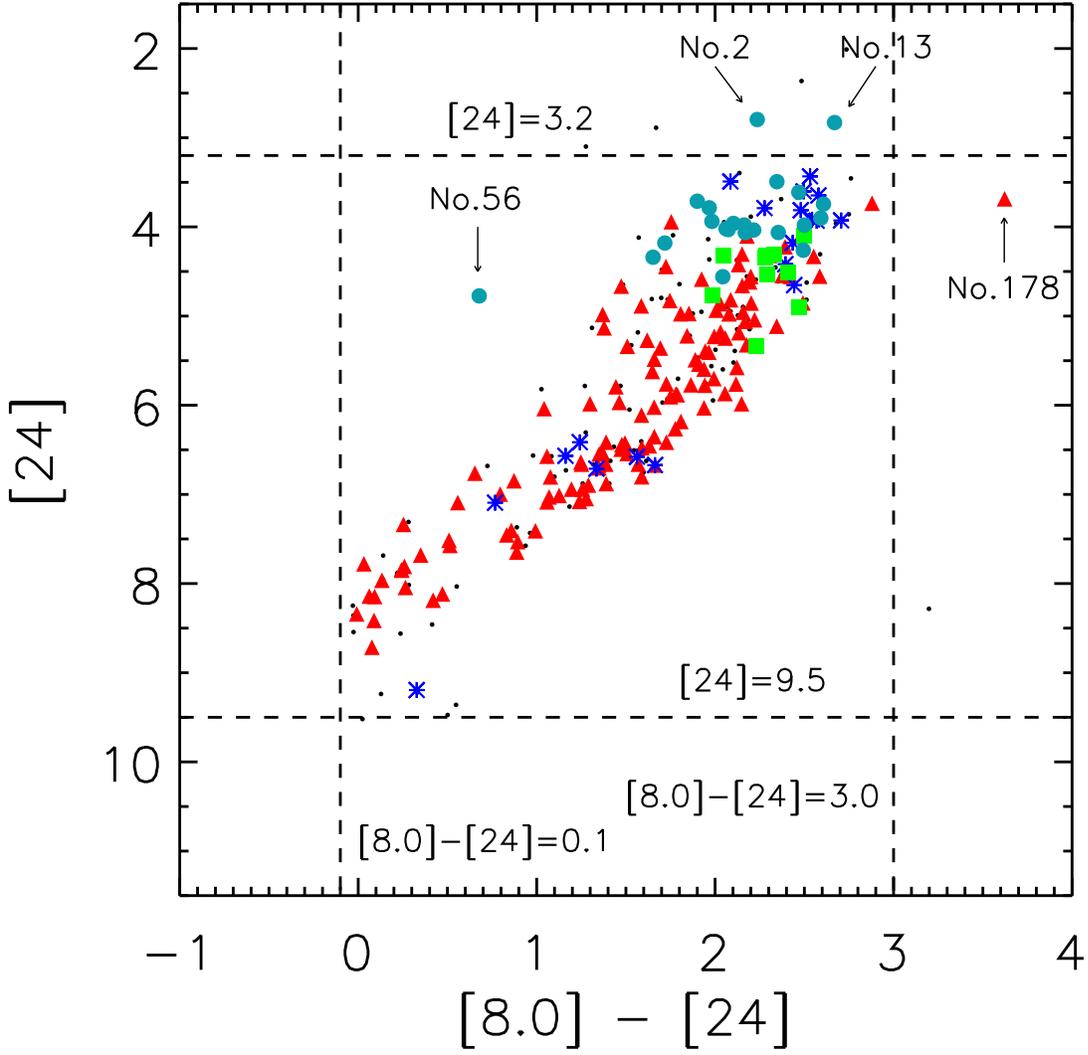}
\caption{
    The same as Fig.~\ref{fig1}, but for [24] vs. [8.0] - [24].  We also
    give the empirical limits for RSGs as
    $3.2 \leq [24] \leq 9.5$ and $-0.1 \leq [8.0] - [24] \leq 3.0$.
} \label{fig3}
\end{figure}

Aside from the [8.0] band, the MIPS [24] band also reflects the
emission of the dust, but the cooler dust \citep{Blum06, Bonanos09}.
Fig.~\ref{fig3} shows the [8.0] - [24]/[24] CMD. No.53 lacks
the [8.0] band data, No.58 lacks the [8.0] and [24] band data,
Nos.8, 73, 75, 93, 117, 147, 161, 167, 169, 182, 184 and 189
lack the [24] band data, they are absent in this diagram. The
targets are divided clearly into two groups. The targets from
\citet{Kastner08}, \citet{Pierce00} and \citet{Feast80} have
redder [8.0]-[24] and thus are RSGs with colder dust, while the
targets from \citet{Massey03} have bluer [8.0]-[24] and are
RSGs with warmer dust. The two groups share similar luminosity
in [24] at the high end. But the redder group do not extend to
the faint end, which is true in [8.0] as well, while not true
in $K_{\rm S}$ since the stellar radiation is the main
contributor in near-infrared. The consistency at the short
wavelengths indicate that the stars of the two groups are more
or less of the same luminosity, while the difference at the
long wavelengths is caused only by the amount and temperature
of dust. So both groups should be RSGs, but with different
dust. As done in previous diagram, we also give the
empirical limits for RSGs as $3.2 \leq [24] \leq 9.5$ and $0.1
\leq [8.0] - [24] \leq 3.0$. Then, there are four outliers.
No.2 and No.13 are a little bit brighter than other targets.
No.56 is bluer and No.178 is redder than the major group. But
since the [24] band mainly reflects the characteristic of
circumstellar dust other than the central star, the variation
of color index should be the influence of dust temperature
which does not directly relate to the central star.

\begin{figure}
\centering
\includegraphics[width=\textwidth, bb=50 280 550 750]{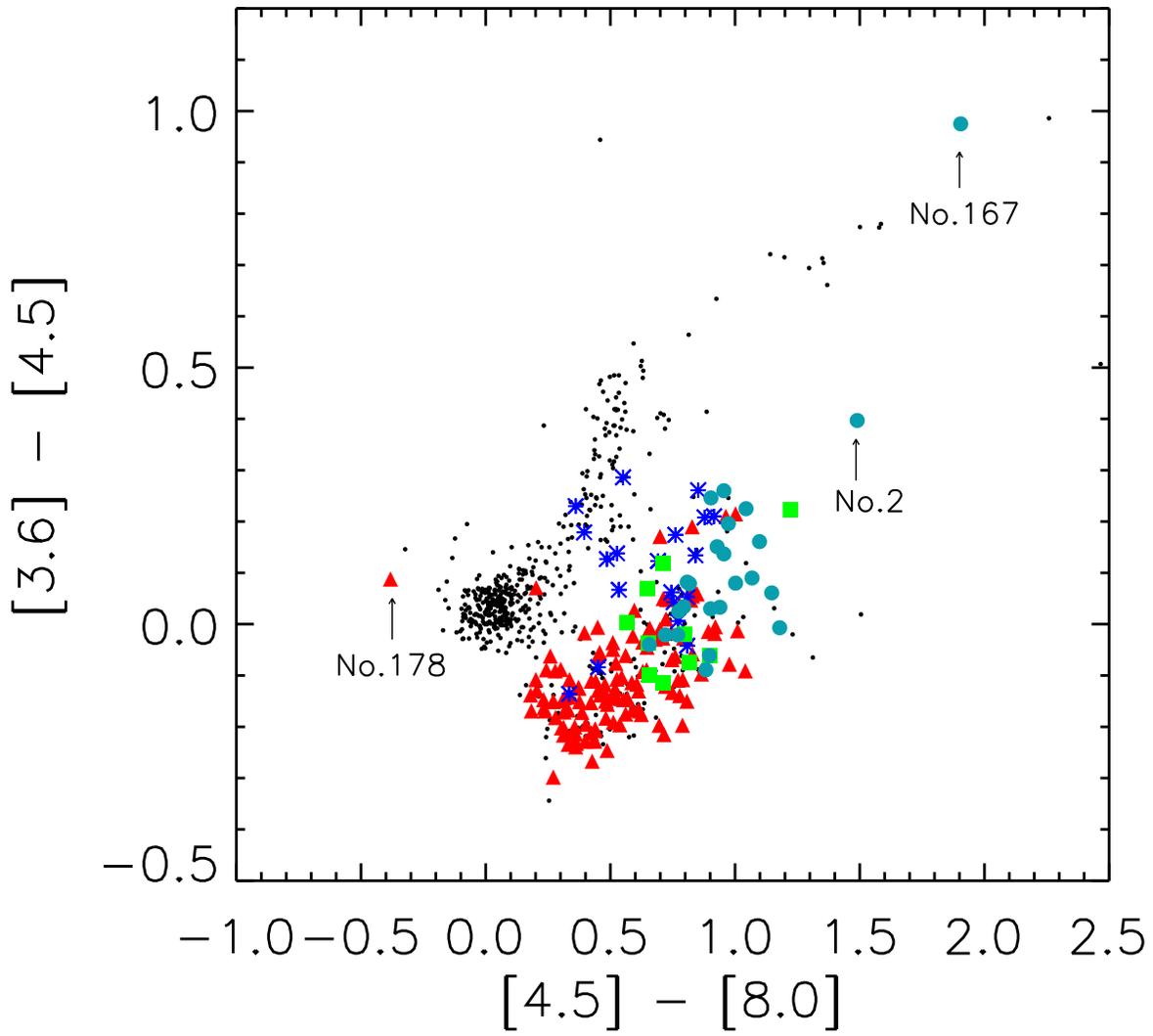}
\caption{
    The color-color diagram [3.6] - [4.5] vs. [4.5] - [8.0]. The symbols have the same meaning as in Fig.~\ref{fig1}.
} \label{fig5}
\end{figure}

\citet{Bonanos09} showed that RSGs locate in a distinct region
in a couple of TCDs. As an example, Fig.~\ref{fig5} is the
[3.6]-[4.5]/[4.5]-[8.0] TCD.  RSGs locate in a region with
 [3.6]-[4.5] bluer than other luminous red stars, which is due
to the depression at [4.5] by the CO bands around 4.6$\mu$m
\citep{Verhoelst09}. Nos.51, 53, 58, 70, 82, 98, 129, and 181
are not present in this diagram due to lack of the measurement
in related bands. There are apparently three outliers, No.2,
No.167 and No.178.

Combining the information from the CMDs and TCDs, we
exclude No.2 and No.167 for their inconsistent colors and
luminosities. A few marginal cases are retained, i.e. No.56
and No.178, because they stay together with the majority of
RSGs in some of the diagrams.

From the CMDs and TCDs, it can be seen that there are about two
dozens massive sources from \citet{Bonanos09} which locate in the
same regions in these diagrams as RSGs but are not included in the sample. These
sources are not included in previous studies like spectroscopy or
detailed analysis, but they may be RSGs judged from their luminosity
and colors. Further observations are needed. This shortage indicates
that the present sample of RSGs is not complete, but seems to
include most RSGs.

\subsection{Period Determination and Analysis}

To collect as long as possible light curves which would make more
accurate period determination, we browsed the online databases of
the ASAS and MACHO projects to search for all useful data. Most of
our photometry data comes from ASAS because RSGs, except several
heavy-envelop-surrounding targets with excess visual extinction,
easily get saturated in the MACHO observation due to their great
intrinsic luminosity \citep{Massey05}. The V magnitude is expected
to be brighter than about 14 mag for RSGs in LMC, estimated from the
minimum mass of RSGs and the distance modulus of LMC, while the
saturation limit is on average about 13 mag in the Kron-Cousins V
band and about 14 mag in the R band (Kem Cook, private
communication) for the MACHO project. Sometimes, an abnormal light
curve is still retrieved from the MACHO database for a target even
it is saturated. As the saturation magnitude depends on the seeing,
the MACHO photometry code SoDOPHOT (basically DoPhot) can measure
more photo-electrons for a stellar image at worse seeing than average
condition. But for this to work, the template image should not be
saturated. The abnormal light curve is indeed caused by the
saturated template image. Thus, all abnormal light curves are
dropped. The photometric precision of MACHO is about 0.02 magnitude,
as an internal error, in both bands of its own two-color system in
field-overlap regions for the brightness between $13\thicksim18$ mag
in the V band \citep{Alcock99}. ASAS has an average precision of
about 0.05 mag, but in some cases (due to problems with
flat-fielding and lack of color information) the errors could be 0.1
magnitude or larger \citep{Pojmanski02}. Although the ASAS
photometric precision is slightly lower than the MACHO project, it
better fits the high luminosity of RSGs and the numerous
observations can compensate partly for the precision. In addition,
by deleting the low-quality data, the ASAS data have reasonably good
precision and time coverage to derive typical optical variation
properties of RSGs.

\begin{figure}
\centering
\includegraphics[width=\textwidth, bb=50 280 550 750]{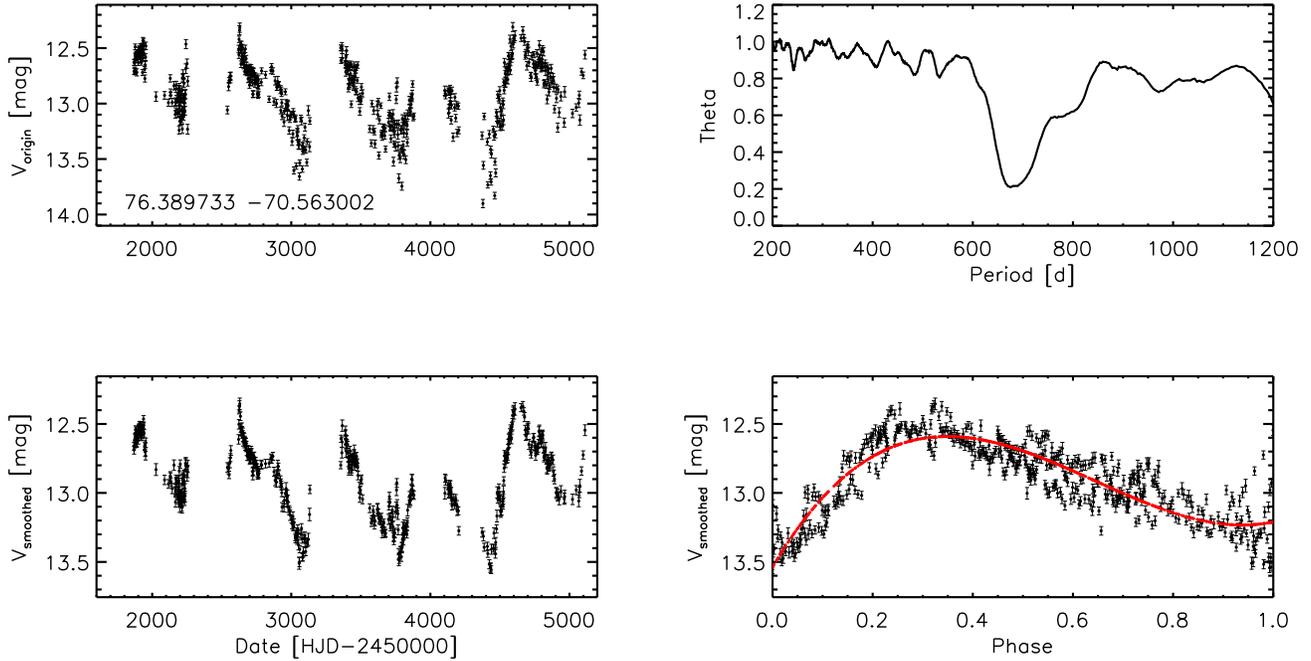}
\caption{
    An example to show the PDM processing of the ASAS data for the star No.34 in Table~\ref{191tab}. Left column from top to bottom: original light curve, smoothed light curve; right column from top to bottom: theta diagram, phase diagram. The coordinates are shown inside the top panel of left column. The red line shows the robust polynomial fitting curve.
} \label{asas_pdm_process}
\end{figure}

The way to handle the photometric data depends on the datasets. For
ASAS, the standard Johnson V band data are used since the released
continuous I band data cover only about 500 days. Because the search
radius is fixed as 30$\arcsec$ to retrieve the photometric data from
the project website, it does need to check the DSS image for the
input coordinates how many stars fall into the search circle. For
the coordinates of each target, the counterpart is accepted only
when its distance to the target is less than 10$\arcsec$
and its average V band magnitude brighter than 14 mag (for a good
S/N to be achieved by ASAS) (Grzegorz Pojmanski, private
communication). The criteria exclude 16 targets with large
coordinate deviations and 14 targets with more than two sources in
the 10$\arcsec$ circle. Moreover, the processed online ASAS data are
graded into four levels marked by A to D that indicate the
photometric quality for each aperture, and we only make use of the A
data with the best quality. We also removed the points which lay
more than $3\sigma$ away from a resistant estimate of the dispersion
of the light curve distribution, where $\sigma$ is the standard
deviation \citep{Hoaglin83}. This selection removes most of
the outlying points. Except some poor photometric measurements,
there are 161 targets with ASAS data useful. Among them, according
to the previous identifications based on the CMDs and TCDs, No.2 has
large coordinate deviation and No.167 has good photometry. After
excluding No.167, 160 targets have ASAS data available. Then the
least-square (Savitzky-Golay) polynomial smoothing filter is applied
to the light curve, which would reduce noise greatly in the
time-series data but retain dynamic range of variations in the data
\citep{Press92}. The process of handling the ASAS photometric data
is shown in the left column of Fig.~\ref{asas_pdm_process}
for one target.

For MACHO, the non-standard two-color photometric system magnitudes
are transformed to the standard Kron-Cousins V and R system by
modifying the formulae of \citet{Alcock99} as $V = V_{M,t} + a0 +
1.089(a1 + 0.022X_{t}) + co + 2.5\log(ET)$ and $R = R_{M,t} + b0 +
1.089(b1 + 0.004X_{t}) + co + 2.5\log(ET)$ in which the arithmetic
average of V-R=1.089 of RSGs is adopted from \citet{Levesque06}. The
search radius is 3$\arcsec$ thanks to the high-accuracy positioning
system and the large aperture of the MACHO telescope. The MACHO
project finally provides the photometric data for 18 targets. Among
them, 9 targets do not have ASAS data due to the coordinate
deviation and the others have useful ASAS photometry data. The
outliers in the CMDs and TCDs in previous section do not have any
MACHO data available. Because of the low temperature and thick
envelope, RSGs are usually quite red and brighter in the R band than
in the V band. They are easily saturated in the R band and left with
only the K-C V band data useful. In the process to convert the
template magnitude to the standard system, the correction for the
air mass, if following the average parameters of the system, would
bring about large uncertainty in the case of large air mass, which
can be understood as the atmosphere changes greatly from night to
night. The data points with airmass larger than 2.0 are thus
deleted, which does not affect the accuracy of the period
significantly as the measurements for every source are numerous.
Besides, the measurement with photometric error bigger than 0.2 mag
is also removed. At last, combining the ASAS and MACHO data, 169
targets have the photometry data  available and useful. The
resource for each target is labeled by 'M' for MACHO and 'A' for
ASAS in Table~\ref{191tab}.

RSGs are found to be variable for long time. But their
variation is not very regular, which makes the period
determination a tough task. On the other hand, an accurate
determination of the period is the key to the period-luminosity
relation. In order to be certain about the period, a couple of
methods are used to obtain the most consistent period.

First, a simple and effective way, the Phase Dispersion
Minimization (PDM) method \citep{Stellingwerf78}, is used to
find the light variation period. In brief, the PDM method
folds the data at a range of trying periods, divides the folded
data into a series of bins and computes the variance of the
amplitude within each bin. The bin variances are combined and
compared to the overall variance of the data set. For a true
period, the ratio of the bin to the total variances, defined as
theta ($\Theta$), should be the minimum and for a false period the
ratio is approximately unity. This method is very useful in
particular for data sets with gaps and non-sinusoidal
variations. Since the light variation of RSGs is mostly
non-sinusoidal, PDM is an appropriate technique. There
is an empirical tip when using this method, i.e., one should
estimate the rough period range via eyes, because the harmonics
of the periods would also produce very small theta, even
smaller than the theta value at the true period. A sample of
the PDM processing is shown in the right column in Fig.~\ref{asas_pdm_process}.
%and Fig.~\ref{macho_pdm_process}.

\begin{figure}
\centering
\includegraphics[width=\textwidth, bb=50 280 550 750]{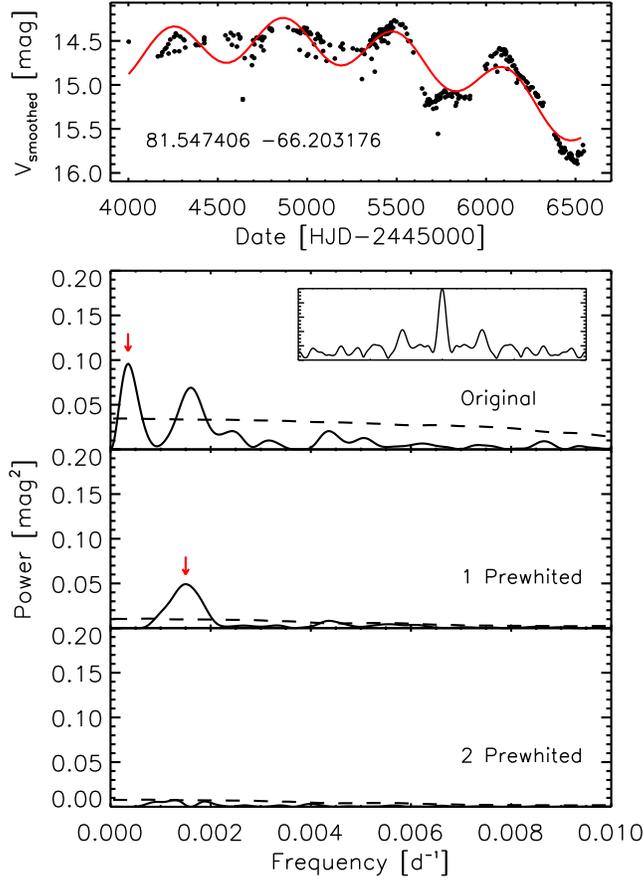}
\caption{
    An example to show the Period04 processing of the MACHO data for the star No.59 in Table~\ref{191tab}. Top panel is the original light curve with red fitting curve; bottom panel is the power spectra. The coordinates are shown inside the top panel. The spectral window in the first power spectrum diagram is shown in the same scale as the power spectra. The dashed line shows the four times noise spectrum. The red arrow marks the highest peak in the power spectra during each iteration.
} \label{asas_p04_process}
\end{figure}

\begin{figure}
\centering
\includegraphics[width=\textwidth, bb=50 280 550 750]{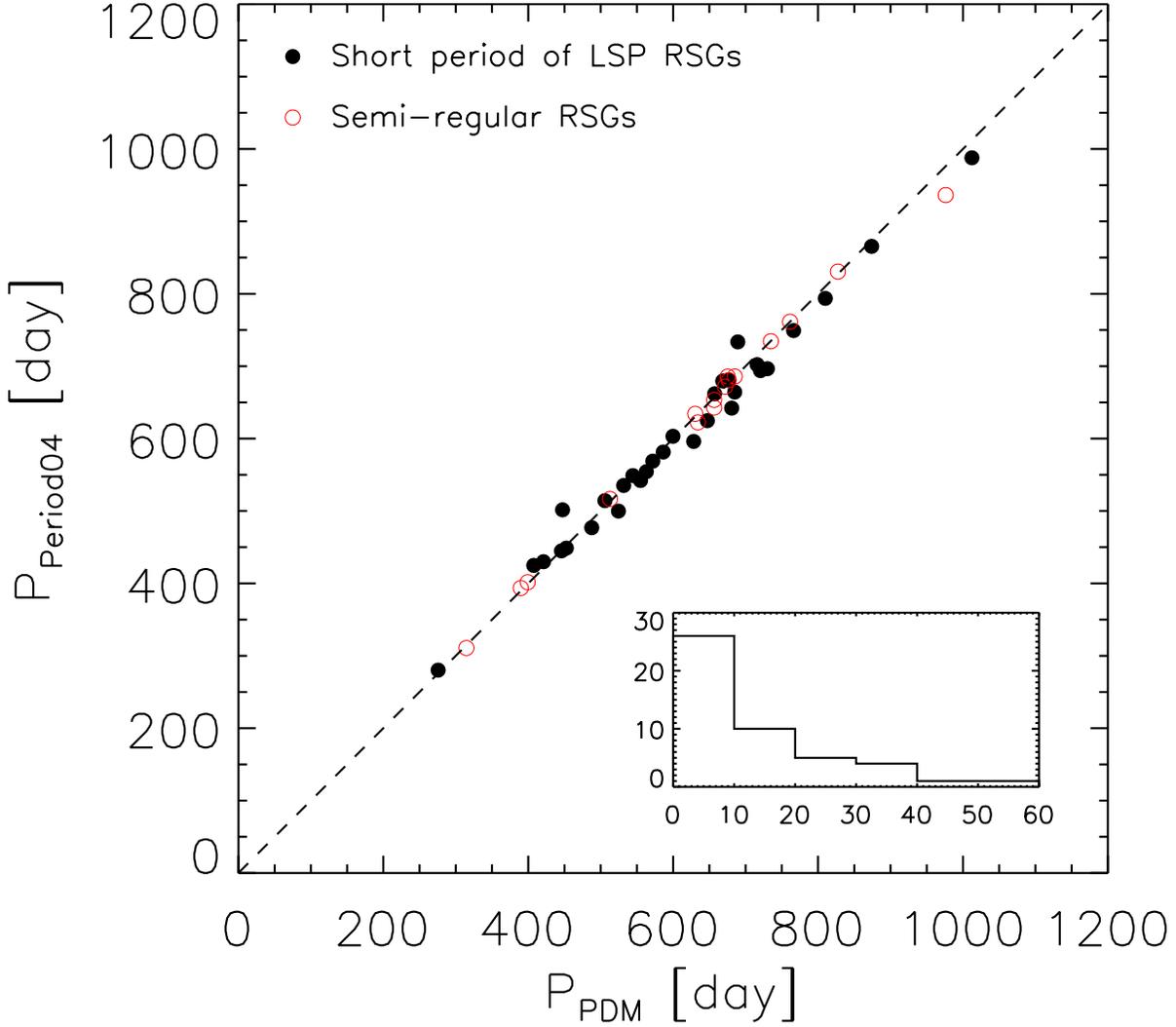}
\caption{
    A comparison of the periods derived from the PDM and Period04 method respectively for the 47 semi-regular RSGs and Distinguishable
    Short Period LSP RSGs. The open red circles indicate the only one period of semi-regular RSGs and the filled black circles indicate the distinguishable short period of LSP RSGs which have both LSP and short period. The inset histogram shows the distribution of the difference between these two methods.
} \label{p04_pdm}
\end{figure}

Although PDM is an outstanding method to detect the light variation
period, there are still some obstacles. A major problem is that our
data has a time span of about 3000 days which is sufficient to
determine the moderately long periods, but not long enough for the
periods over 1500 days because the time coverage is less than two
periods. Actually, Long Secondary Periods (LSPs) which are found in
the AGBs and RGs variables are also present in RSGs. The LSPs are
often thousands of days long, sometimes exceeding the time span
of the data set and leading to an unreliable period determination
by PDM. In such case, the Period04 \citep{Lenz04} method based on
the Fourier transform works better. So, in addition to the PDM
method, Period04 is also used to analyze the variability of the
targets. One purpose of using Period04 is to extract the long LSP in
some cases, the other is to confirm the period derived from PDM. For
the later purpose, Period04 implements iterative sinusoid fitting to
fit and subtract a sinusoid match with the frequency at the highest
peak in power spectrum in each iteration. After first iteration, the
residual data are used to calculate the power spectrum in the
following iterations. The iteration is stopped until the highest peak
in the residual spectrum is less than four times of the noise level.
A sample of Period04 processing is shown in
Fig.~\ref{asas_p04_process}. Only when the periods derived from PDM
and Period04 agree with each other, is the period regarded as true.
Fig.~\ref{p04_pdm} shows a comparison of the periods derived from
PDM and Period04, and the inset is the histogram of the difference
between these two periods for 47 RSGs. It can be seen that the
difference is mostly less than 10 days. From the consistency between
PDM and Period04, the period is regarded to be real.

\section{Period-Luminosity Relation}

\begin{figure}
\centering
\includegraphics[width=\textwidth, bb=20 530 555 666]{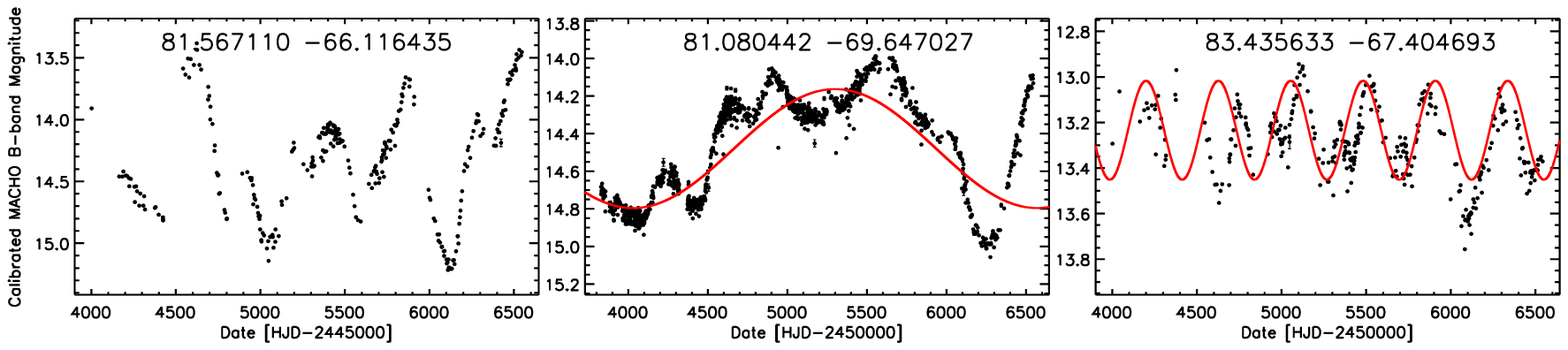}
\includegraphics[width=\textwidth, bb=20 530 555 666]{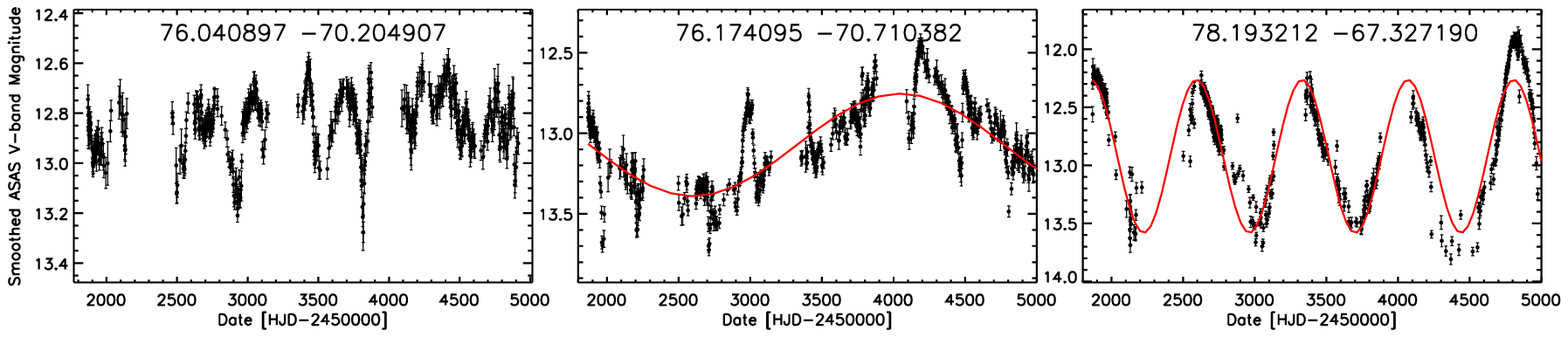}
\caption{
     The sample light curves of three types of variation.
     From left to right: irregular light curve,
    long secondary period light curve and semi-regular light curve. Top panel: the ASAS data,
    bottom panel: the MACHO data. In each diagram, the coordinates
    are shown in the top. Red line shows the fitted curve. }
\label{lightcurve}
\end{figure}

According to the analysis of the time-series photometric data, the
targets are divided into five categories. The first category
includes 20 RSGs. They are either too bright and saturated in
photometry or in a crowded stellar field or having a close companion
impossible to resolve. These stars have poor photometry data and are
not considered for any further analysis. The
second category includes 35 RSGs which have complex lightcurve not
suitable for deriving the P-L relation. The third category
includes 23 RSGs. They are irregular variables, for which there is
no possibility to find an appropriate period to characterize its
light variation. They are neither considered for further study of
the P-L relation in the following part of this paper. But they are
the characters in our next paper on the period change of RSGs. The remaining 111
RSGs are semi-regular or LSP RSGs which are involved in the
following determination of the P-L relation and classified into the
fourth and fifth categories. The fourth category includes 16 RSGs,
being semi-regular variables with period statistical significance
less than or equal to 0.05. Among them, 4 targets are in
\citet{Feast80}, 1 in \citet{Pierce00}, 4 in \citet{Kastner08} and 7
in \citet{Massey03}. One thing to keep in mind is that there is no
clear borderline between semi-regular and irregular variables. What
we do is to calculate the statistical significance of the period
corresponding to the minimum theta and classify the object as a
semi-regular variable when the significance is smaller than 0.05
\citep{Schwarzenberg97}. The fifth category includes 95 RSGs which
have LSP, 31 of them have distinguishable short period. For the 16 semi-regular variable
RSGs and those 31 LSP RSGs but with distinguishable short period,
the period can be determined with relatively high accuracy, and
suitable for discussing the P-L relation. Their periods are derived
and the results are shown in Table \ref{srpdmtab}. For an intuitive
view, we give the irregular, semi-regular and LSP RSGs each a sample
light curve in Fig.~\ref{lightcurve}.

\begin{figure}
\centering
\includegraphics[width=\textwidth, bb=50 280 550 750]{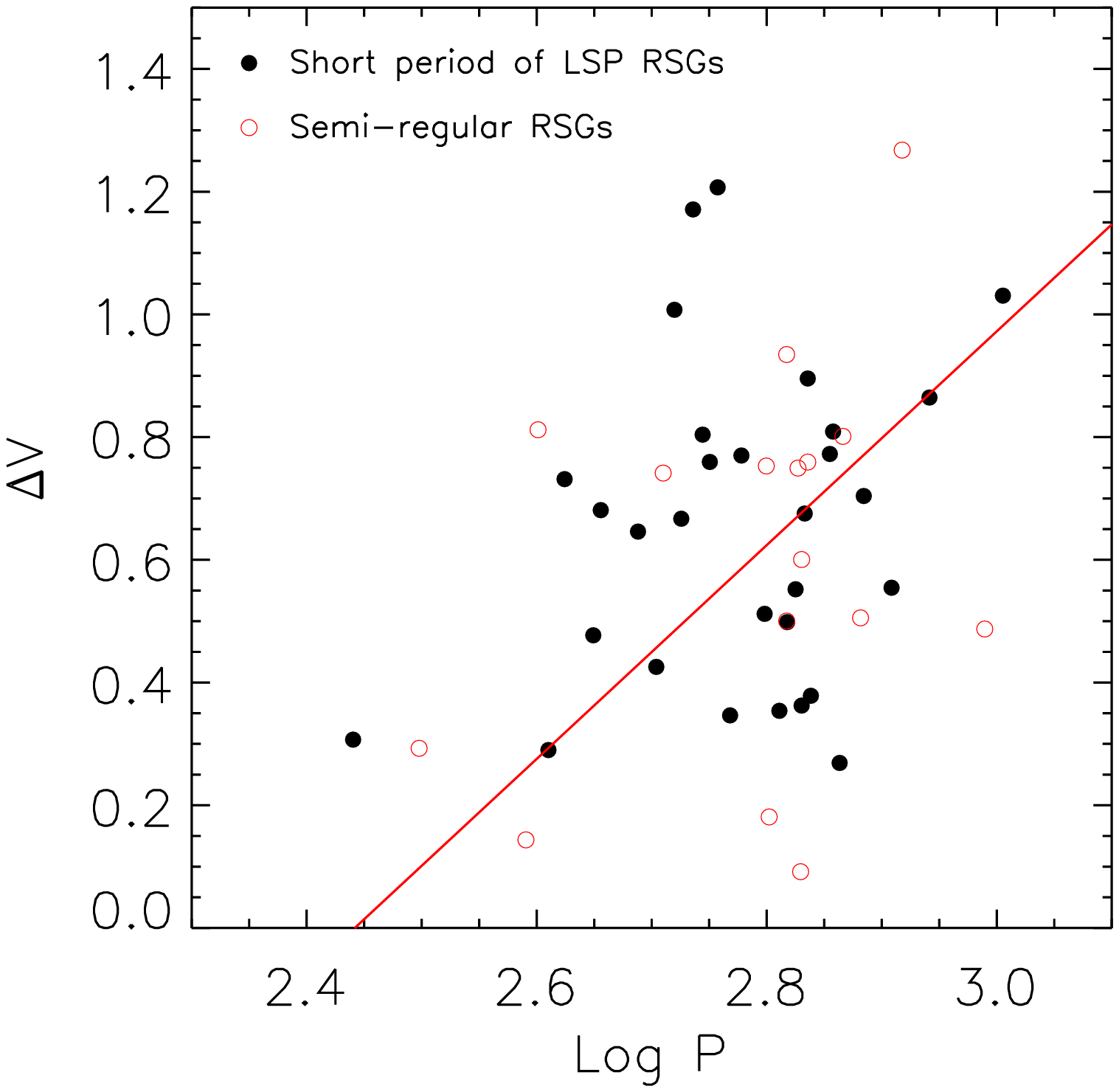}
\caption{
    The period and the full amplitude of the 47 semi-regular RSGs  and  LSP RSGs with distinguishable short
    period. The red solid line is a linear fit between the amplitude
    and the period.
} \label{fig9}
\end{figure}

We also calculate the linear relation of amplitude with period for the RSGs
present in Table \ref{srpdmtab}. The result is
$\bigtriangleup V = (1.74 \pm 0.54) \times {\rm P} - (4.25 \pm
1.13)$, with rms=0.74, where $\bigtriangleup V$ is the full
amplitude in the $V$ band and $P$ is the period. This is consistent
with the general tendency of variables that the longer the period
the greater the amplitude. The distribution of the full amplitude
with the period is shown in Fig.~\ref{fig9}. The average period
$\bar{P} = 618$ day and the average amplitude in the V band
$\bigtriangleup \bar{V} = 0.698$mag.

Beside the period, the other key parameter in the P-L relation is the luminosity. Usually
the luminosity is derived from the brightness in one specific band
by converting through the bolometric correction. The band often used
is the visual V band or the near-infrared K band.  Such method has
to suffer the bolometric correction factor that is relatively
uncertain for RSGs. The V band further suffers the interstellar
extinction from both the Galaxy and the LMC, and even the
circumstellar dust, which may be serious and more importantly is inhomogeneous. The choice of the
indicator of the luminosity has to be cautious.

\begin{figure}
\centering
\includegraphics[width=\textwidth, bb=50 280 550 750]{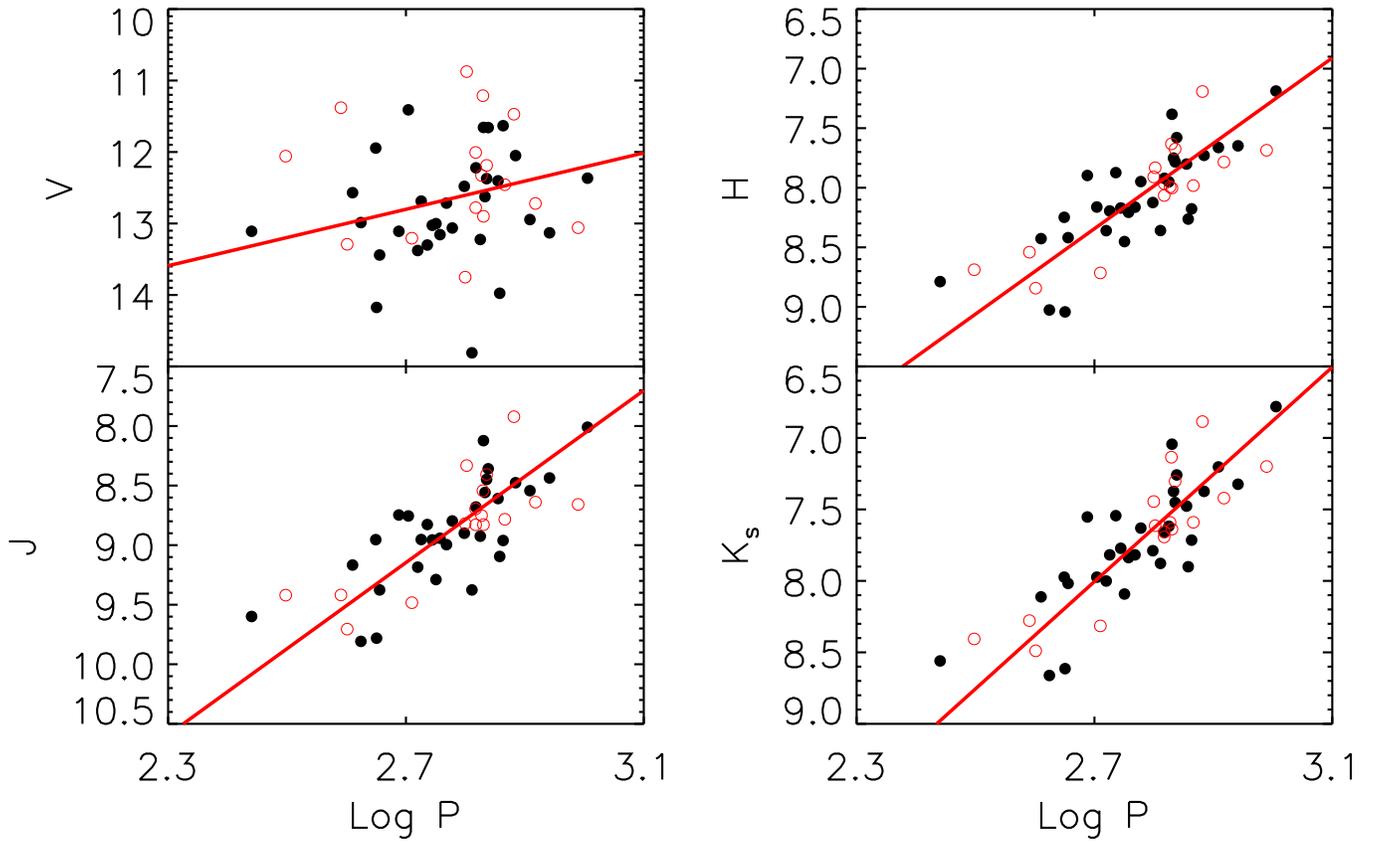}
\caption{
   The Period-Luminosity relation in the V, J, H, and K$_{S}$ bands
   for the 47 semi-regular RSGs and  LSP RSGs with distinguishable short
    period. The symbols are the same as in Fig.~\ref{fig9}.
} \label{pl-semi-int1}
\end{figure}

\begin{figure}
\centering
\includegraphics[width=\textwidth, bb=50 280 550 750]{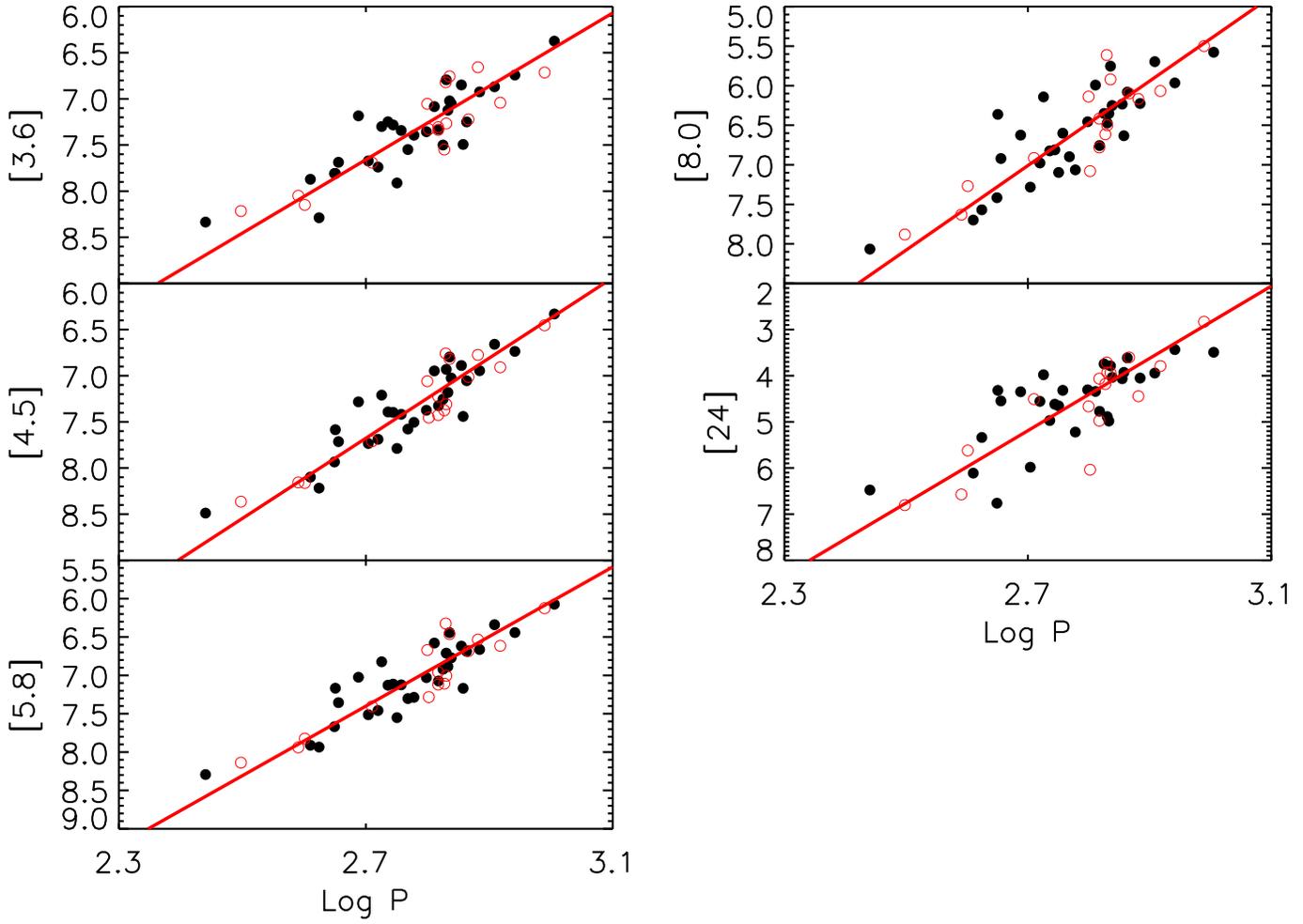}
\caption{
    The same as Fig.~\ref{pl-semi-int1}, but the P-L relation is in the \emph{Spitzer}/IRAC [3.6], [4.5], [5.8], [8.0] and \emph{Spitzer}/MIPS [24] bands. } \label{pl-semi-int2}
\end{figure}

As a trying, the period is plotted against the brightness in
various bands, the visual band V, the near-infrared band $K_{s}$ and
the \emph{Spitzer}/IRAC bands. The V band completely reflects the
radiation of the stellar photosphere, but it can be seriously
affected by extinction (A$_{\rm V}$=0.6 mag at E(B-V)=0.2 for
the Galactic extinction at the direction to LMC), in particular
the targets should experience inhomogeneous extinction because they
locate in different environments in LMC, which can be seen in
Fig.~\ref{lmc}. Even if we correct the foreground Galactic
interstellar extinction, it is impossible to correct the
extinction of LMC since we are neither clear about the
extinction structure of LMC nor the depth of the targets. So no
interstellar extinction is corrected in any band, this for sure
underestimates the luminosity, much in visual bands, a little in
near-infrared bands, and little in \emph{Spitzer}/IRAC bands.
Meanwhile, the brightness in the
\emph{Spitzer}/IRAC long-wavelength band [8.0] must be at least partly from the
emission of the surrounded dust so that it is not a right
indicator of the stellar luminosity, but it has the advantage
of best avoiding the effect of interstellar extinction. In
spite of such various shortcomings, the linear analysis of the P-L
relation is performed to every band involved in the way
$M_{\lambda} = a \log P + b$. The slope $a$ and intercept $b$
with their dispersions are listed in Table~\ref{pltab}, and the
fitted lines are shown in Fig.~\ref{pl-semi-int1} and
Fig.~\ref{pl-semi-int2}.

The period-luminosity relationship does exist in the semi-regular
RSG variables and the LSP with distinguishable short period in all
bands, but the amount of dispersion depends on the band. In the V
band the dispersion is the biggest. In the J band, and the MIPS [24]
and [8.0] bands, the dispersion is relatively big, but the P-L
relations are reasonably good. In the H, K, [3.6] and [5.8] bands,
the dispersion is very small, and the P-L relation is very tight.
The best relationship occurs in the [4.5] band. However, with very
small difference, the later two groups, i.e. H, K, [3.6], [4.5] and
[5.8], all have a highly reliable P-L relation.  This can be
understood. In these bands, their brightness should come mainly from
the stellar photosphere if the targets are cold and have no thick
dust envelope. Besides, the extinction is very small in these bands.
According to the newest estimation of extinction in the IRAC bands,
the extinction in the [3.6] and [4.5] bands is only 63\% and 57\% of
the K band extinction \citep{Gao09}, which means only about 0.04 mag
at E(B-V)=0.2 for the Galactic forground extinction, significantly smaller than the
photometric error. Moreover,  the variation amplitude in the
infrared should be very small from the decreasing tendency of the
amplitude with the wavelength, so that the one-epoch magnitude can
represent the average brightness very approximately. Therefore, with the
large-scale data available from \emph{Spitzer}, the short IRAC bands
are recommended as the luminosity indicator.

\begin{figure}
\centering
\includegraphics[width=\textwidth, bb=50 280 550 750]{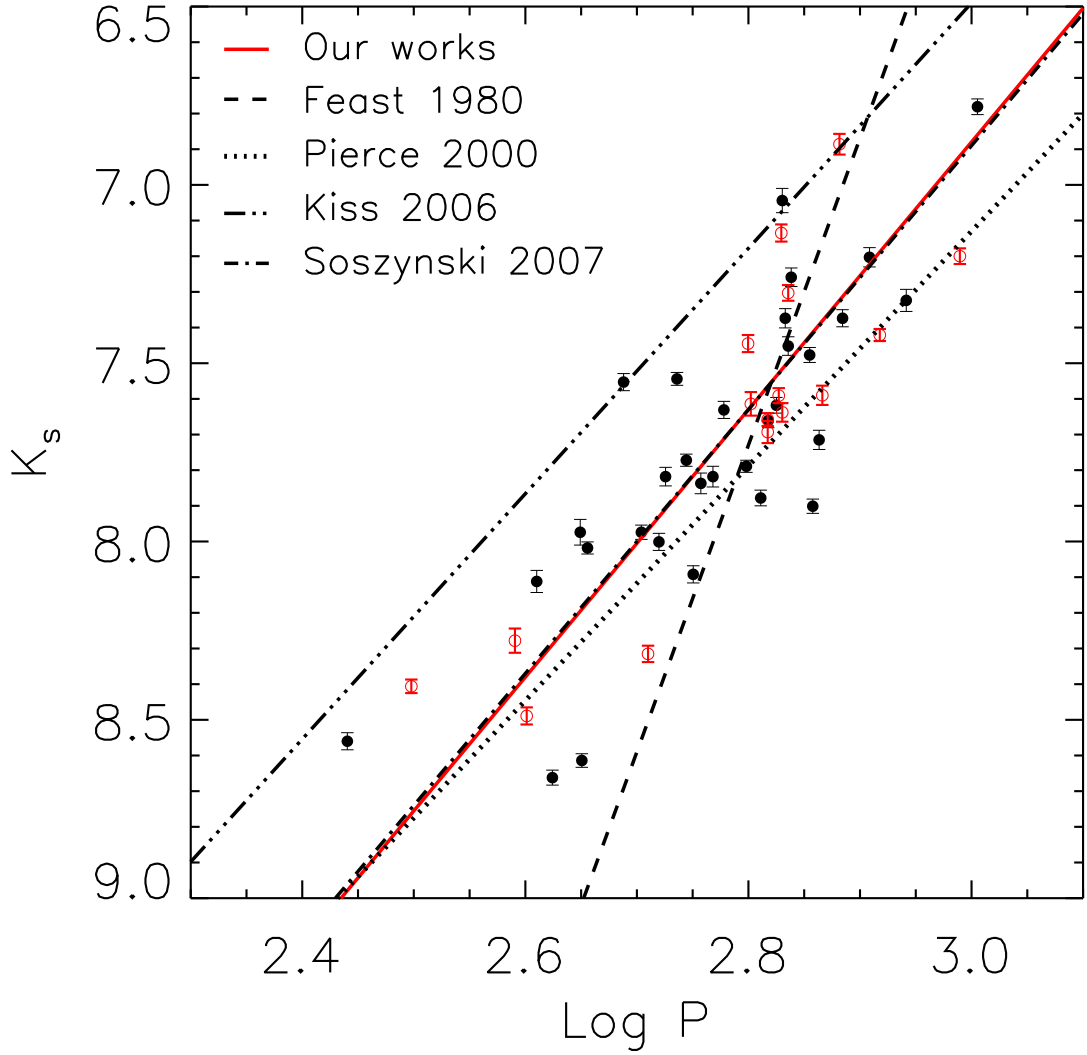}
\caption{ The P-L relation in the $K_{\rm S}$ band. The symbols have
the same meaning as in Fig.~\ref{pl-semi-int1}, together with the
measurement error bar in the $K_{\rm S}$ band. Also shown are the
P-L relations for RSGs obtained by \citet{Feast80}, \citet{Pierce00}
    and \citet{Kiss06}. The  $a2$ sequence for the AGB stars
    from \citet{Soszynski07aca} is almost superposed on ours.
} \label{pl-semi-k}
\end{figure}

In the $K_{\rm S}$ band, previous investigators already obtained the
P-L relation for RSGs. In Fig.~\ref{pl-semi-k}, our result
is compared with the P-L relation for RSGs in LMC by \citet{Feast80}
and in the Per OB1 association, LMC and M33 by \citet{Pierce00}.
It can be seen that our result doesn't agree perfectly with
either of them, but close to that of \citet{Pierce00}, and very
different from that of \citet{Feast80}. Our result is also
compared with the P-L relation for RSGs in our Galaxy by \citet{Kiss06}. Our result brings about
systematically fainter luminosity than \citet{Kiss06}, indeed, the
\citet{Kiss06} relation produces higher luminosity at a given
period than that of \citet{Feast80} and \citet{Pierce00} as
well. However, our P-L relation is almost perfectly matched
with the extension of the AGB $a_{2}$ sequence of
\citet{Soszynski07aca}, where the objects are the so-called
OGLE Small Amplitude Red Giants (OSARGs), corresponding to the
sequence B of \citet{Wood99} and identified as the first overtone
radial pulsating red giants. But whether these RSGs are
pulsating in the first overtone mode needs further proof because
no other sequences such as the fundamental and second overtone modes
are accompanied.

As the LSP is also a periodic phenomenon, the LSP RSGs are checked
whether the P-L relation exists. The periods of LSP RSGs are
calculated from the power density spectrum using the Period04 code
\citep{Lenz04}. The periods and amplitudes in the V-band of
these RSGs are shown in Table \ref{lsptab}, where only 51 targets
with the period shorter than 3000 days are listed, while the other 44 RSGs have  very possibly a longer period exceeding the length of the data set that makes the period determination uncertain.
It can be told that they have periods of a
few thousand days, much longer than that of the semi-regular RSGs,
expected from their identity as Long Secondary Period variables.
Besides, the amplitude of variation is only a few tenths magnitude,
much smaller than that of the semi-regular RSGs. Both the period and
amplitude indicate that the LSP RSGs are a different type of
variables from the semi-regular RSGs. Their origin of variation
should be different too. Indeed, the origin of LSP in
semi-regular variable Miras is very controversial. Many models,
including radial and non-radial pulsation, binary, stellar spot
models, are suggested but none of them receives general acceptance
\citep{Wood09, Nie10}. For the LSP in RSGs, it has not even
attracted much attention except for \citet{Kiss06}.  The data
covering  much longer time are the key to study the behavior of the
LSP of RSGs. With the continuation of the ASAS project, a more
reliable determination of the period and analysis of the mechanism are
promising.

\begin{figure}
\centering
\includegraphics[width=\textwidth, bb=40 290 500 720]{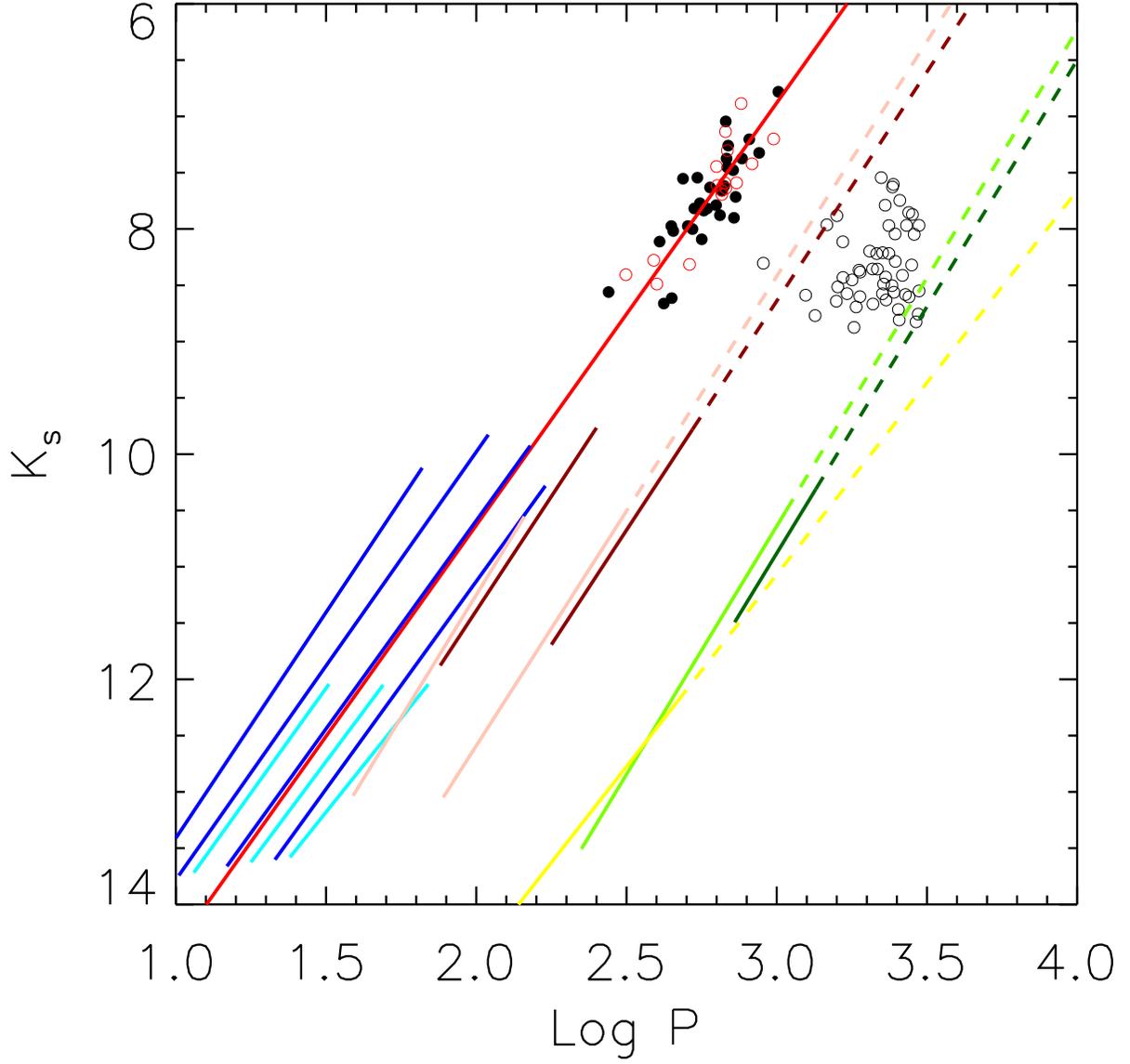}
\caption{
    The $K_{\rm S}$-band P-L relation for the semi-regular RSGs and
    LSP RSGs superposed on the P-L relations of LPVs in the LMC by \citet{Soszynski07aca}.
    The symbols obey the same convention as previous except that the newly added black open circle denotes the long secondary period of the LSP RSGs.
    The \citet{Soszynski07aca} lines are the same as their Fig.~2. }
\label{pl-semi+lsp}
\end{figure}

For an overview of the P-L relation of the semi-regular and the
LSP RSGs, Fig.~\ref{pl-semi+lsp} shows their period and brightness
in the $K_{s}$ band in comparison with the P-L relations of red giants
by \citet{Soszynski07aca}. It again exhibits an almost perfect
match with the extension of the Sosyznski $a_{2}$ sequence, and the
LSP lies between the C and D sequences. Different from the red giants,
the RSGs do not form well-defined sequences. The technical reason
may be the smallness of the sample and the relatively short duration of the time
series. But the irregularity of the variation
of RSGs brings about the difficulty in determining period and the
dispersion of the relation as well.

\section{Summary}

The P-L relation of RSGs has been known to exist that gives an
opportunity to extend the distance calibration from these
luminous and reasonably numerous targets. With the latest optical
and infrared survey data, we obtained a series of
period-luminosity relations in several infrared bands, new or
different from previous studies.

The preliminary sample of RSGs is ever the largest by combining a
few  systematic samples. To obtain a reliable match,
the infrared data of our sample are selected by strict
criterion with a one-arcsec search radius with the
\emph{Spitzer}/IRAC and MIPS databases. Although this criterion
drops some targets with a little big position error, the loss
in the size of the sample is acceptable from the original large
sample. The final sample consists of 191 sources. To further
purify the sample to be consisted of only RSGs, two more targets
are dropped because they have slightly different colors and luminosity
from the bulky members according to their locations in a few
CMDs and TCDs.

The time-series photometric data are mainly taken from the ASAS
project, supplemented by the MACHO project. Usually more than
300 measurements are available for one target and the time span is longer than
a few thousand days. The period analysis is performed carefully
by the PDM method and further checked by  the Period04 code for consistency. Based on
the derived period and amplitude as well as the shape of light
curve, the RSGs are classified into five categories: 20 RSGs
saturated in photometry or located in crowded stellar field, 35
RSGs with complex light curve, 23 RSGs with irregular variation,
16 RSGs with semi-regular variation, 95 RSGs with LSPs among
which 31 have distinguishable short period, and 51 have
a long period shorter than 3000 days that can be determined with reasonable accuracy while the remaining 44 objects have a long period exceeding the length of the data set.

The P-L relation is found to exist both in the semi-regular
variable and LSP RSGs with distinguishable short period. This relationship is analyzed in various
bands, from visual $V$, through near-infrared $JHK_{\rm S}$, to
the mid-infrared \emph{Spitzer}/IRAC/MIPS bands. Except the V band, the P-L relation is tight in all the other bands, and  it has the least dispersion in the IRAC [3.6] and [4.5]
bands that are recommended for use.

For the derived P-L relation, some attentions must be paid.
First, the sample has 191 RSGs, but only 47 obey the P-L
relation, i.e. 24\%. It means that many RSGs do not obey the
P-L relation. Before making use of the P-L relation, for
example as a distance calibrator for distant galaxies, one
must make sure that the tracer RSG has an accurately determined
period. Moreover, the P-L relation is tighter at longer wavelengths, 3.6
and 4.5 micron in our cases, which is not affected
significantly by not only interstellar extinction but also the
surrounded dust of RSG itself. For this purpose, the photometry
in such bands are needed.

Our work is roughly consistent with \citet{Pierce00} in the
K-band P-L relation but with better precision at longer
wavelengths. \citet{Pierce00} gave a simultaneous fitting for
RSGs in the Per OB1 association of the Galaxy, LMC and M33
which may participate more uncertainty in this relation. The
difference with \citet{Kiss06} could be due to the difference
in the metallicity between Galaxy and LMC or the methods in
period determination.

\section{Acknowledgements}

We thank Prof. P. Wood for very helpful discussion, and the
anonymous referee for useful suggestions. This work is
supported by China's NSFC through the projects 10778601 and
10973004, China 973 Program 2007CB815406, and the Fundamental
Research Funds for the Central Universities.

\clearpage

%%%%%%%%%%%%%%%%%%%%%%%%%%%%%%%%%%%%%%%%%%%%%%%%%%%%%%%%%%%%%%%%%%%%%%%%%%%%%%%%%%

\begin{deluxetable}{ccccccccccccc}
\tabletypesize{\scriptsize}
\tablecaption{
        Infrared brightness of the 191 RSG candidates
    \label{191tab}
}
\tablewidth{0pt}
\tablehead{
\colhead{No.} & \colhead{RA (${\degr}$)} & \colhead{Decl (${\degr}$)} &
\colhead{J} & \colhead{H} & \colhead{K$_{\rm S}$} &
\colhead{[3.6]} & \colhead{[4.5]} & \colhead{[5.8]} &
\colhead{[8.0]} & \colhead{[24]} &\colhead{Data\tablenotemark{a}} &\colhead{Reference\tablenotemark{b}}
}
\startdata
1  &  72.343634&  -69.409571&    9.052&   8.171&  7.763&  7.376&  7.297&  7.036&  6.480&  3.980&      A&  K\\
2  &  72.422868&  -68.630874&    9.130&   8.047&  7.486&  6.919&  6.522&  6.096&  5.032&  2.796&  -----&  K\\
3  &  72.744524&  -69.234115&    9.433&   8.603&  8.355&  8.008&  8.247&  8.000&  7.886&  6.638&      A&  M\\
4  &  72.879183&  -69.247765&    9.426&   8.549&  8.239&  7.951&  8.165&  7.919&  7.804&  6.415&      A&  M\\
5  &  72.947048&  -69.323501&    9.863&   9.033&  8.740&  8.539&  8.711&  8.526&  8.325&  7.047&      A&  M\\
6  &  73.311619&  -69.204979&    9.143&   8.134&  7.751&  7.514&  7.563&  7.299&  7.053&  4.975&      A&  M\\
7  &  73.326893&  -69.284163&    9.517&   8.670&  8.356&  8.098&  8.274&  8.020&  7.650&  7.092&      A&  M\\
8  &  73.378763&  -69.297135&    9.321&   8.457&  8.056&  7.491&  7.458&  7.117&  6.519&  -----&      A&  K\\
9  &  73.653572&  -69.339499&    8.332&   7.832&  7.614&  7.329&  7.454&  7.283&  7.080&  6.038&      A&  M\\
10 &  73.660602&  -69.188074&    8.543&   7.662&  7.203&  6.869&  6.659&  6.340&  5.696&  3.942&      A&  M\\
11 &  73.664248&  -69.076768&    9.808&   9.026&  8.662&  8.288&  8.218&  7.935&  7.569&  5.338&      A&  P\\
12 &  73.707049&  -69.500735&    9.484&   8.728&  8.429&  8.122&  8.177&  7.939&  7.720&  6.847&      A&  M\\
13 &  73.762753&  -69.486868&    8.658&   7.685&  7.200&  6.714&  6.454&  6.125&  5.499&  2.830&      A&  K\\
14 &  73.816926&  -69.320038&    8.558&   7.751&  7.374&  7.123&  7.181&  6.885&  6.352&  4.982&      A&  M\\
15 &  73.840164&  -69.787996&    8.925&   7.950&  7.618&  7.500&  7.254&  6.919&  6.351&  3.744&      A&  K\\
16 &  73.875008&  -69.486257&    8.696&   7.965&  7.658&  7.337&  7.426&  7.119&  6.781&  4.974&      A&  M\\
17 &  73.883554&  -66.843868&    8.680&   7.921&  7.661&  7.329&  7.325&  7.073&  6.757&  4.772&      A&  P\\
18 &  73.895278&  -69.448796&    8.233&   7.448&  7.113&  6.857&  6.946&  6.581&  6.063&  3.961&      A&  K\\
19 &  73.924266&  -69.440054&    8.800&   8.026&  7.695&  7.449&  7.485&  7.218&  6.831&  4.537&      A&  P\\
20 &  73.951116&  -69.401813&    9.127&   8.280&  7.968&  7.710&  7.925&  7.618&  7.210&  5.180&      A&  M\\
21 &  74.098594&  -69.703102&    9.512&   8.701&  8.454&  8.232&  8.448&  8.266&  8.136&  6.941&      A&  M\\
22 &  74.117827&  -69.676949&    9.504&   8.657&  8.427&  8.148&  8.330&  8.184&  8.050&  6.702&      A&  M\\
23 &  74.381387&  -70.149824&    9.157&   8.291&  7.973&  7.474&  7.356&  7.074&  6.645&  4.313&      A&  P\\
24 &  74.430466&  -70.147335&    8.436&   7.648&  7.324&  6.742&  6.736&  6.442&  5.965&  3.433&      A&  F\\
25 &  74.435802&  -69.509524&    9.694&   8.849&  8.562&  8.278&  8.489&  8.275&  8.142&  7.084&      A&  M\\
26 &  75.539736&  -70.417168&    9.465&   8.613&  8.322&  8.077&  8.220&  8.050&  7.904&  6.535&      A&  M\\
27 &  75.813714&  -70.294932&    9.782&   8.985&  8.715&  8.637&  8.725&  8.536&  8.424&  7.530&      A&  M\\
28 &  76.020954&  -70.379563&    9.167&   8.427&  8.112&  7.871&  8.099&  7.911&  7.698&  6.111&      A&  M\\
29 &  76.040897&  -70.204907&    9.465&   8.676&  8.390&  8.207&  8.408&  8.191&  8.096&  7.963&      A&  M\\
30 &  76.058805&  -67.270660&    8.010&   7.188&  6.781&  6.374&  6.331&  6.073&  5.577&  3.491&      A&  F\\
31 &  76.174095&  -70.710382&    9.184&   8.380&  8.029&  7.817&  7.893&  7.619&  7.371&  5.406&      A&  M\\
32 &  76.225852&  -70.555205&    9.493&   8.669&  8.404&  8.252&  8.419&  8.219&  8.088&  7.845&      A&  M\\
33 &  76.291645&  -70.667688&    9.429&   8.647&  8.383&  8.215&  8.365&  8.170&  8.094&  7.850&      A&  M\\
34 &  76.389733&  -70.563002&    8.828&   8.002&  7.638&  7.267&  7.309&  7.005&  6.501&  3.932&      A&  F\\
35 &  76.486300&  -70.589963&    9.234&   8.440&  8.114&  7.791&  7.860&  7.542&  7.113&  4.957&      A&  M\\
36 &  76.495647&  -70.487290&    9.569&   8.806&  8.472&  8.334&  8.457&  8.230&  8.085&  6.454&      A&  M\\
37 &  76.498117&  -70.803169&    9.483&   8.715&  8.315&  7.691&  7.711&  7.405&  6.914&  4.507&      A&  P\\
38 &  76.651730&  -70.544063&    9.811&   9.022&  8.750&  8.474&  8.611&  8.438&  8.131&  5.983&  -----&  M\\
39 &  76.773682&  -70.545627&    8.123&   7.383&  7.044&  6.793&  6.931&  6.710&  6.472&  4.886&      A&  M\\
40 &  76.885651&  -70.651219&    9.376&   8.419&  8.018&  7.687&  7.714&  7.355&  6.921&  4.545&      A&  M\\
41 &  77.367843&  -68.797713&   10.045&   9.095&  8.658&  8.270&  8.143&  7.937&  7.657&  6.415&      M&  F\\
42 &  77.431671&  -65.366457&    8.829&   8.065&  7.693&  7.307&  7.225&  6.958&  6.417&  4.063&      A&  K\\
43 &  78.193212&  -67.327190&    8.783&   7.982&  7.590&  7.223&  7.013&  6.683&  6.098&  3.602&      A&  F\\
44 &  78.707198&  -67.455496&    8.639&   7.784&  7.421&  7.042&  6.908&  6.615&  6.067&  3.790&      A&  F\\
45 &  79.287434&  -69.539210&    9.015&   8.163&  7.824&  7.517&  7.444&  7.177&  6.615&  4.223&      A&  M\\
46 &  79.484786&  -69.673710&    9.702&   8.870&  8.575&  8.387&  8.587&  8.395&  8.230&  6.661&      A&  M\\
47 &  79.763600&  -69.665335&    8.530&   7.913&  7.604&  7.379&  7.332&  7.027&  6.510&  4.586&      A&  M\\
48 &  79.972044&  -69.459329&    9.250&   8.447&  8.218&  7.823&  7.962&  7.726&  7.447&  5.539&      A&  M\\
49 &  80.098385&  -69.557466&    9.080&   8.329&  7.983&  7.766&  7.711&  7.427&  6.882&  4.330&      A&  M\\
50 &  80.366502&  -69.504510&    9.311&   8.504&  8.145&  8.046&  7.987&  7.696&  7.139&  4.553&      A&  M\\
51 &  80.629539&  -69.568088&    9.784&   8.963&  8.649&  -----&  8.510&  8.209&  7.793&  6.997&      A&  M\\
52 &  80.761457&  -69.343640&    9.625&   8.827&  8.513&  8.261&  8.473&  8.280&  8.049&  6.542&      A&  M\\
53 &  80.891611&  -69.318588&    -----&   -----&  -----&  -----&  8.851&  8.737&  -----&  8.144&      A&  M\\
54 &  80.931710&  -65.699946&    9.011&   8.131&  7.744&  7.321&  7.287&  7.020&  6.493&  3.901&      A&  K\\
55 &  80.975513&  -70.168343&   10.039&   9.119&  8.774&  8.741&  8.674&  8.431&  8.140&  6.578&      M&  M\\
56 &  81.080442&  -69.647027&    8.360&   7.375&  6.809&  6.384&  6.354&  6.061&  5.452&  4.774&    A M&  K\\
57 &  81.436850&  -69.080232&    9.178&   8.336&  7.993&  7.566&  7.675&  7.362&  6.904&  4.817&      A&  M\\
58 &  81.499584&  -69.565144&   11.718&  10.961& 10.731& 10.597& 10.426& 10.257&  -----&  -----&      M&  F\\
59 &  81.547406&  -66.203176&    9.376&   8.359&  7.878&  7.084&  6.947&  6.579&  5.992&  4.340&    A M&  K\\
60 &  81.567110&  -66.116435&    9.187&   8.375&  8.027&  7.835&  7.684&  7.309&  6.756&  4.262&    A M&  K\\
61 &  81.614147&  -69.182168&    8.934&   8.078&  7.705&  7.430&  7.491&  7.158&  6.594&  4.096&      A&  P\\
62 &  81.617605&  -69.132679&    9.606&   8.759&  8.483&  8.246&  8.450&  8.237&  8.011&  6.352&      A&  M\\
63 &  81.645022&  -68.861091&    8.490&   7.599&  7.265&  6.831&  6.852&  6.563&  6.082&  4.021&  -----&  K\\
64 &  81.675405&  -68.944066&    9.696&   8.817&  8.550&  8.330&  8.537&  8.255&  8.099&  7.029&      A&  M\\
65 &  81.678049&  -68.953632&    9.613&   8.826&  8.550&  8.148&  8.156&  7.888&  7.498&  5.322&      A&  M\\
66 &  81.792519&  -69.607455&   10.399&   9.544&  9.159&  8.942&  8.829&  8.501&  8.232&  6.99 &      M&  M\\
67 &  81.792920&  -69.271529&    9.812&   9.002&  8.784&  8.533&  8.669&  8.505&  8.334&  6.671&      A&  F\\
68 &  81.809231&  -69.186349&    9.355&   8.551&  8.192&  7.870&  7.893&  7.607&  7.303&  5.246&      A&  M\\
69 &  81.861454&  -69.521000&    9.727&   8.942&  8.666&  8.516&  8.666&  8.437&  8.185&  6.895&      A&  M\\
70 &  81.866884&  -69.010018&    9.405&   8.592&  8.323&  8.187&  -----&  7.988&  7.679&  6.021&      A&  M\\
71 &  81.873700&  -67.236976&    9.030&   8.298&  7.971&  7.528&  7.338&  6.980&  6.510&  5.132&      A&  M\\
72 &  81.893100&  -66.891674&    8.942&   8.206&  7.837&  7.343&  7.418&  7.123&  6.601&  4.316&      A&  P\\
73 &  81.915221&  -69.150377&    9.056&   8.253&  7.970&  7.707&  7.941&  7.700&  7.610&  -----&      A&  M\\
74 &  81.947921&  -69.222399&    8.815&   7.998&  7.604&  7.275&  7.282&  6.857&  6.103&  4.030&      A&  K\\
75 &  81.962977&  -67.301129&    9.704&   8.904&  8.601&  8.375&  8.642&  8.380&  8.215&  -----&      A&  M\\
76 &  81.963061&  -69.179399&    9.445&   8.626&  8.288&  8.100&  8.263&  8.075&  8.030&  7.680&      A&  M\\
77 &  82.024881&  -69.120357&    9.416&   8.578&  8.149&  7.584&  7.536&  7.273&  6.810&  4.658&      A&  M\\
78 &  82.033607&  -69.219738&    9.185&   8.360&  8.001&  7.739&  7.689&  7.458&  6.977&  4.555&      A&  M\\
79 &  82.064241&  -66.981324&    9.289&   8.452&  8.092&  7.911&  7.788&  7.551&  7.097&  4.654&      A&  F\\
80 &  82.066198&  -69.200268&    9.625&   8.800&  8.514&  8.281&  8.579&  8.360&  8.308&  8.043&      A&  M\\
81 &  82.077459&  -69.126368&    9.486&   8.657&  8.314&  8.115&  8.282&  8.038&  7.670&  5.882&      A&  M\\
82 &  82.116319&  -69.215940&    8.974&   8.666&  8.384&  -----&  7.934&  7.739&  7.431&  5.968&      A&  M\\
83 &  82.120235&  -68.118893&    8.609&   7.801&  7.477&  6.849&  6.889&  6.619&  6.233&  4.064&      A&  K\\
84 &  82.126448&  -69.012338&    9.653&   8.772&  8.503&  8.393&  8.429&  8.203&  7.919&  6.424&      A&  M\\
85 &  82.131423&  -69.091968&    9.290&   8.396&  8.049&  7.802&  7.932&  7.652&  7.319&  5.186&      A&  M\\
86 &  82.177000&  -69.128438&    9.988&   9.286&  8.995&  8.613&  8.623&  8.358&  7.924&  5.869&  -----&  M\\
87 &  82.179938&  -67.307893&    9.613&   8.822&  8.582&  8.327&  8.573&  8.308&  8.086&  7.572&      A&  M\\
88 &  82.189510&  -68.967304&    8.710&   7.941&  7.549&  7.295&  7.464&  7.185&  6.847&  5.339&      A&  M\\
89 &  82.215933&  -70.012402&   10.244&   9.346&  8.980&  8.712&  8.574&  8.333&  8.047&  6.712&      M&  F\\
90 &  82.249936&  -67.750382&   11.482&  10.556& 10.066&  9.889&  9.973&  9.724&  9.522&  9.194&      M&  F\\
91 &  82.253219&  -68.775960&    9.557&   8.739&  8.435&  8.188&  8.286&  8.008&  7.635&  5.771&      A&  M\\
92 &  82.264496&  -69.112835&    9.095&   8.263&  7.901&  7.494&  7.441&  7.168&  6.633&  3.927&      A&  F\\
93 &  82.272933&  -67.304874&    9.695&   8.847&  8.574&  8.466&  8.615&  8.442&  8.307&  -----&      A&  M\\
94 &  82.285018&  -69.205096&    9.458&   8.668&  8.345&  8.084&  8.259&  8.016&  7.698&  5.576&      A&  M\\
95 &  82.337464&  -68.792041&    8.954&   8.248&  7.974&  7.809&  7.935&  7.669&  7.416&  6.762&      A&  M\\
96 &  82.339290&  -69.005629&    8.833&   8.033&  7.746&  7.340&  7.437&  7.052&  6.572&  4.826&      A&  M\\
97 &  82.364952&  -69.147313&    8.407&   7.675&  7.303&  6.756&  6.817&  6.463&  5.920&  3.938&      A&  K\\
98 &  82.393450&  -66.924538&    9.846&   9.051&  8.730&  -----&  7.800&  7.490&  6.812&  4.418&    A M&  F\\
99 &  82.425819&  -68.954832&    7.922&   7.192&  6.886&  6.657&  6.774&  6.534&  6.168&  4.444&      A&  M\\
100&  82.433175&  -69.097210&    8.989&   8.180&  7.882&  7.563&  7.536&  7.288&  6.940&  4.934&      A&  M\\
101&  82.478150&  -69.071027&    9.420&   8.688&  8.406&  8.215&  8.364&  8.138&  7.881&  6.804&      A&  M\\
102&  82.478929&  -67.310215&    8.900&   8.124&  7.789&  7.356&  7.374&  7.029&  6.456&  4.305&      A&  M\\
103&  82.509512&  -67.045854&    8.755&   8.162&  7.974&  7.672&  7.734&  7.513&  7.282&  5.983&      A&  M\\
104&  82.519134&  -68.791330&    9.804&   9.032&  8.769&  8.574&  8.801&  8.596&  8.401&  7.408&      A&  M\\
105&  82.520562&  -69.066599&    9.900&   9.075&  8.808&  8.590&  8.781&  8.546&  8.269&  6.879&      A&  M\\
106&  82.539901&  -69.184357&    9.839&   8.942&  8.641&  8.391&  8.619&  8.402&  8.261&  7.403&      A&  M\\
107&  82.587314&  -67.334840&    8.448&   7.783&  7.452&  7.022&  6.797&  6.444&  5.753&  3.786&      A&  K\\
108&  82.592061&  -67.108752&    9.697&   8.932&  8.603&  8.438&  8.589&  8.441&  8.318&  7.078&      A&  M\\
109&  82.609471&  -69.506780&    9.598&   8.822&  8.478&  8.285&  8.432&  8.187&  7.880&  5.763&      A&  M\\
110&  82.639484&  -67.287532&    9.215&   8.393&  8.121&  7.998&  8.089&  7.934&  7.810&  7.778&      A&  M\\
111&  82.648062&  -68.989827&    8.747&   7.897&  7.553&  7.184&  7.283&  7.024&  6.625&  4.347&      A&  P\\
112&  82.648212&  -67.201204&    9.913&   9.159&  8.852&  8.726&  8.868&  8.720&  8.535&  7.647&      A&  M\\
113&  82.672671&  -69.259404&    8.752&   7.993&  7.591&  7.549&  7.375&  7.108&  6.614&  4.179&      A&  F\\
114&  82.674892&  -69.089771&    9.889&   9.079&  8.756&  8.464&  8.630&  8.381&  8.040&  6.263&      A&  M\\
115&  82.688204&  -67.133163&    9.598&   8.781&  8.431&  8.174&  8.402&  8.148&  7.966&  6.493&      A&  M\\
116&  82.717866&  -67.292882&    9.865&   9.040&  8.827&  8.618&  8.726&  8.566&  8.390&  6.802&      A&  M\\
117&  82.751940&  -69.177779&   10.728&   9.849&  9.554&  9.298&  9.467&  9.283&  9.284&  -----&      A&  M\\
118&  82.755027&  -69.183132&    9.480&   8.623&  8.328&  8.020&  8.100&  7.739&  7.223&  5.228&      A&  M\\
119&  82.764302&  -69.094463&    9.621&   8.787&  8.582&  8.375&  8.513&  8.366&  8.333&  8.341&      A&  M\\
120&  82.767398&  -69.317501&    9.022&   8.060&  7.627&  6.981&  6.901&  6.519&  5.899&  4.181&      A&  K\\
121&  82.788629&  -67.431940&    8.797&   7.948&  7.631&  7.394&  7.505&  7.286&  7.064&  5.221&      A&  M\\
122&  82.814424&  -69.066357&    9.394&   8.524&  8.219&  7.919&  8.027&  7.741&  7.237&  5.793&      A&  M\\
123&  82.826852&  -69.157832&    9.778&   8.954&  8.628&  8.434&  8.496&  8.352&  8.237&  8.145&      A&  M\\
124&  82.903446&  -66.502146&    8.476&   7.728&  7.374&  6.924&  6.945&  6.663&  6.225&  4.050&      A&  K\\
125&  82.947519&  -67.384246&    9.594&   8.842&  8.588&  8.350&  8.468&  8.231&  7.990&  6.182&      A&  M\\
126&  83.036714&  -67.188513&    9.764&   8.975&  8.692&  8.405&  8.600&  8.408&  8.198&  6.941&      A&  M\\
127&  83.080233&  -67.416724&    9.914&   9.144&  8.874&  8.692&  8.839&  8.734&  8.607&  8.187&      A&  M\\
128&  83.114295&  -69.281282&    9.075&   8.261&  7.963&  7.766&  7.858&  7.595&  7.228&  5.059&      A&  M\\
129&  83.130593&  -69.340385&    9.744&   8.901&  8.630&  -----&  8.670&  8.420&  8.286&  7.454&      A&  M\\
130&  83.147117&  -69.131018&    9.466&   8.575&  8.263&  7.975&  7.981&  7.817&  7.532&  5.594&      A&  M\\
131&  83.209336&  -67.462504&    9.179&   8.331&  8.048&  7.734&  7.931&  7.581&  7.142&  5.483&      A&  M\\
132&  83.281690&  -66.801580&    9.781&   9.042&  8.614&  7.808&  7.585&  7.167&  6.363&  4.317&    A M&  P\\
133&  83.310319&  -67.063479&    9.432&   8.587&  8.320&  8.051&  8.283&  8.045&  7.919&  6.442&      A&  M\\
134&  83.361725&  -67.070424&    8.953&   8.195&  7.818&  7.300&  7.210&  6.823&  6.142&  3.979&      A&  K\\
135&  83.373315&  -67.527127&    9.949&   9.151&  8.825&  8.682&  8.790&  8.656&  8.589&  8.117&      A&  M\\
136&  83.435633&  -67.404693&    9.705&   8.843&  8.489&  8.148&  8.162&  7.824&  7.269&  5.621&    A M&  M\\
137&  83.467379&  -69.187088&    9.220&   8.356&  7.899&  7.440&  7.401&  7.090&  6.624&  4.320&      A&  M\\
138&  83.558570&  -68.978860&    9.272&   8.370&  7.956&  7.574&  7.313&  6.996&  6.461&  3.918&    A M&  F\\
139&  83.581189&  -68.993541&    9.634&   8.822&  8.533&  8.349&  8.493&  8.250&  7.967&  6.029&      A&  M\\
140&  83.589252&  -69.366740&    9.912&   9.113&  8.891&  8.775&  8.704&  8.589&  8.502&  8.413&      A&  M\\
141&  83.640732&  -69.250679&    9.465&   8.580&  8.278&  8.017&  8.200&  8.003&  7.718&  5.776&      A&  M\\
142&  83.695939&  -69.483476&   10.068&   9.275&  9.020&  8.867&  8.996&  8.877&  8.791&  8.714&      A&  M\\
143&  83.808775&  -67.732191&    9.376&   8.533&  8.024&  7.379&  7.171&  6.840&  6.293&  3.813&      A&  F\\
144&  83.828775&  -67.038816&    9.541&   8.706&  8.336&  7.970&  8.084&  7.783&  7.373&  4.902&      A&  P\\
145&  83.852212&  -69.067616&    9.567&   8.657&  8.234&  7.785&  7.801&  7.495&  7.056&  4.854&    A M&  M\\
146&  83.867980&  -66.934027&    8.359&   7.579&  7.259&  7.050&  7.025&  6.771&  6.251&  4.035&      A&  K\\
147&  83.886724&  -69.071993&    9.449&   8.567&  8.197&  7.934&  8.105&  7.909&  7.791&  -----&      A&  M\\
148&  83.932594&  -68.855835&    9.121&   8.270&  8.044&  7.690&  7.829&  7.511&  7.051&  5.358&      A&  M\\
149&  83.966537&  -69.374751&    9.567&   8.706&  8.451&  8.271&  8.440&  8.263&  8.202&  8.140&      A&  M\\
150&  83.980150&  -69.166502&    8.543&   7.632&  7.135&  6.820&  6.759&  6.324&  5.612&  3.712&      A&  K\\
151&  84.026578&  -68.944660&    9.548&   8.738&  8.442&  8.168&  8.370&  8.195&  8.066&  7.807&      A&  M\\
152&  84.044369&  -68.911174&    9.164&   8.319&  7.968&  7.453&  7.531&  7.161&  6.554&  4.423&      A&  M\\
153&  84.084997&  -68.938532&    9.704&   8.743&  8.354&  7.840&  7.931&  7.426&  6.890&  5.271&    A M&  M\\
154&  84.106074&  -66.927346&    8.638&   7.888&  7.496&  7.033&  6.971&  6.712&  6.226&  3.647&      A&  F\\
155&  84.111546&  -69.397579&    8.482&   8.039&  7.811&  7.611&  7.624&  7.218&  6.614&  3.735&      A&  M\\
156&  84.169085&  -69.387874&    9.872&   9.061&  8.812&  8.499&  8.544&  8.268&  7.897&  6.547&      A&  M\\
157&  84.335449&  -69.327410&    9.418&   8.541&  8.278&  8.049&  8.156&  7.939&  7.630&  6.572&      A&  M\\
158&  84.359859&  -68.794516&    9.406&   8.576&  8.226&  7.976&  8.105&  7.863&  7.654&  5.904&      A&  M\\
159&  84.377738&  -69.042524&    9.928&   9.057&  8.721&  8.450&  8.561&  8.361&  8.138&  7.012&      A&  M\\
160&  84.403673&  -69.489890&    9.455&   8.560&  8.211&  8.028&  8.055&  7.830&  7.345&  4.853&      A&  M\\
161&  84.429627&  -69.416735&    9.058&   8.188&  7.871&  7.626&  7.776&  7.473&  6.969&  -----&      A&  M\\
162&  84.437941&  -69.346846&    8.961&   8.177&  7.715&  7.250&  7.054&  6.687&  6.081&  3.612&      A&  K\\
163&  84.494492&  -69.239996&    9.532&   8.716&  8.383&  7.969&  7.986&  7.781&  7.590&  7.336&      A&  M\\
164&  84.527712&  -69.291588&    9.339&   8.395&  7.913&  7.489&  7.610&  7.283&  6.889&  4.861&      A&  M\\
165&  84.566732&  -69.169779&    9.308&   8.336&  7.867&  7.498&  7.283&  6.925&  6.281&  4.102&  -----&  M\\
166&  84.575514&  -69.295149&    9.607&   8.647&  8.304&  7.994&  8.127&  7.733&  7.379&  5.490&      A&  M\\
167&  84.641794&  -69.342182&   10.211&   9.379&  8.508&  7.251&  6.276&  5.559&  4.371&  -----&      A&  K\\
168&  84.942562&  -69.324538&    9.698&   8.761&  8.473&  8.193&  8.390&  8.105&  7.694&  5.701&      A&  M\\
169&  84.987101&  -69.589185&   10.759&   9.900&  9.589&  9.540&  9.361&  9.138&  8.966&  -----&      M&  F\\
170&  85.032042&  -69.334746&    9.481&   8.610&  8.286&  8.054&  8.089&  7.831&  7.459&  5.114&  -----&  M\\
171&  85.070984&  -69.465015&    9.373&   8.487&  8.220&  7.995&  7.985&  7.675&  7.262&  5.041&      A&  M\\
172&  85.102157&  -69.354771&    9.140&   8.259&  7.854&  7.539&  7.560&  7.244&  6.755&  4.555&      A&  M\\
173&  85.105752&  -69.258410&    9.774&   9.073&  8.783&  8.525&  8.354&  8.079&  7.655&  5.874&      A&  M\\
174&  85.154144&  -69.439026&    9.494&   8.674&  8.322&  8.034&  8.097&  7.775&  7.337&  5.392&      A&  M\\
175&  85.182553&  -69.366179&    8.819&   7.909&  7.445&  7.054&  7.059&  6.670&  6.138&  4.664&      A&  M\\
176&  85.202334&  -69.560030&    9.165&   8.201&  7.711&  7.098&  6.937&  6.552&  5.839&  3.493&      A&  K\\
177&  85.230778&  -69.390385&    8.827&   7.873&  7.544&  7.248&  7.392&  7.128&  6.824&  4.971&      A&  M\\
178&  85.246957&  -69.310076&    -----&   -----&  -----&  7.013&  6.925&  6.652&  7.307&  3.685&      A&  M\\
179&  85.271164&  -69.078427&    9.230&   8.319&  7.975&  7.820&  7.976&  7.723&  7.488&  5.761&      A&  M\\
180&  85.278948&  -69.287421&    8.956&   8.170&  7.772&  7.283&  7.397&  7.114&  6.811&  4.617&      A&  M\\
181&  85.294802&  -69.634524&    8.789&   7.982&  7.631&  -----&  7.274&  6.951&  6.602&  4.559&      A&  K\\
182&  85.340759&  -69.530296&    8.996&   8.163&  7.818&  7.549&  7.578&  7.302&  6.898&  -----&      A&  M\\
183&  85.373065&  -69.454438&    9.589&   8.727&  8.452&  8.249&  8.400&  8.208&  8.024&  7.515&      A&  M\\
184&  85.430854&  -69.470982&    9.520&   8.716&  8.412&  8.172&  8.275&  8.061&  7.730&  -----&      A&  M\\
185&  85.433494&  -69.200771&    9.579&   8.704&  8.400&  8.122&  8.348&  8.107&  7.912&  6.661&      A&  M\\
186&  85.458991&  -69.354324&    9.599&   8.788&  8.560&  8.335&  8.488&  8.293&  8.066&  6.477&      A&  M\\
187&  85.503057&  -69.193601&    9.900&   9.017&  8.683&  8.390&  8.586&  8.345&  8.048&  6.661&  -----&  M\\
188&  85.660761&  -69.164291&    9.930&   9.119&  8.816&  8.642&  8.703&  8.455&  8.142&  6.415&      A&  M\\
189&  85.758491&  -69.097160&    9.502&   8.637&  8.366&  8.172&  8.261&  8.129&  8.018&  -----&      A&  M\\
190&  87.305605&  -70.711291&   10.823&  10.019&  9.445&  8.567&  8.281&  8.060&  7.730&  6.568&      M&  F\\
191&  88.116003&  -69.236126&   10.605&   9.754&  9.204&  8.451&  8.221&  8.062&  7.860&  7.093&      M&  F\\
\enddata
\tablenotetext{a}{Photometry data resource: 'A' for ASAS, 'M' for MACHO and '---' for null. }
\tablenotetext{b}{ Reference: 'F' for \citet{Feast80}, 'P' for \citet{Pierce00}, 'M' for \citet{Massey03}, 'K' for \citet{Kastner08}}
\end{deluxetable}

\clearpage

%%%%%%%%%%%%%%%%%%%%%%%%%%%%%%%%%%%%%%%%%%%%%%%%%%%%%%%%%%%%%%%%%%%%%%%%%%%%%%%%%%%%%%%%%%%%%%%%%%%

%%%%%%%%%%%%%%%%%%%%%%%%%%%%%%%%%%%%%%%%%%%%%%%%%%%%%%%%%%%%%%%%%%%%%%%%%%%%%%%%%%%%%%%%%%%%%%%%%%%

\begin{deluxetable}{ccccccccc}
%\rotate
\tabletypesize{\scriptsize}
\tablecaption{
        Results of the visual light variation by the PDM and Period04 methods for the 47 semi-regular and LSP RSGs with
        distinguishable short period
    \label{srpdmtab}
}
\tablewidth{0pt}
\tablecolumns{9}
\tablehead{
    \colhead{} &
    \colhead{} &
    \multicolumn{3}{c}{PDM} &
    &
    \multicolumn{2}{c}{Period04} \\ \cline{3-5} \cline{7-8}
    \colhead{No.} &
    \colhead{$\langle$$m_{\rm V}$$\rangle$} &
    \colhead{Period (d)} &
    \colhead{Amplitude (mag)} &
    \colhead{Theta} &
    &
    \colhead{Period (d)} &
    \colhead{Amplitude (mag)} &
    \colhead{Measurements}
}
\startdata
9  &  10.87&    634&     0.18&   0.51&&   625&   0.07&    478  \\
10 &  12.94&    810&     0.55&   0.49&&   800&   0.20&    532  \\
11 &  12.98&    420&     0.73&   0.68&&   431&   0.24&    440  \\
13 &  13.05&    976&     0.48&   0.45&&   943&   0.17&    254  \\
14 &  12.62&    680&     0.67&   0.41&&   645&   0.27&    529  \\
15 &  13.22&    668&     0.55&   0.34&&   680&   0.24&    337  \\
16 &  12.00&    656&     0.50&   0.43&&   653&   0.22&    500  \\
17 &  12.22&    657&     0.49&   0.40&&   662&   0.20&    511  \\
24 &  13.13&    873&     0.86&   0.30&&   869&   0.60&    493  \\
28 &  12.56&    407&     0.29&   0.51&&   425&   0.08&    518  \\
30 &  12.36&   1012&     1.03&   0.29&&   990&   0.68&    585  \\
34 &  12.90&    676&     0.60&   0.21&&   684&   0.32&    513  \\
37 &  13.20&    512&     0.74&   0.41&&   518&   0.38&    549  \\
39 &  11.65&    676&     0.36&   0.45&&   684&   0.14&    561  \\
40 &  13.44&    452&     0.68&   0.60&&   450&   0.23&    544  \\
42 &  12.77&    656&     0.93&   0.42&&   645&   0.48&    510  \\
43 &  12.45&    734&     0.80&   0.14&&   735&   0.49&    520  \\
44 &  12.72&    827&     1.26&   0.12&&   833&   0.83&    817  \\
59 &  14.81&    647&     1.63&   0.49&&   625&   0.25&    296  \\
72 &  13.15&    571&     1.20&   0.41&&   571&   0.66&    888  \\
78 &  13.37&    524&     1.00&   0.57&&   500&   0.35&    508  \\
79 &  13.00&    562&     0.75&   0.50&&   555&   0.27&    981  \\
83 &  12.40&    715&     0.77&   0.39&&   704&   0.31&    816  \\
92 &  13.97&    720&     0.80&   0.44&&   694&   0.33&    396  \\
95 &  11.94&    445&     0.47&   0.50&&   446&   0.22&    934  \\
97 &  12.18&    684&     0.75&   0.30&&   689&   0.34&    624  \\
99 &  11.47&    761&     0.50&   0.26&&   763&   0.28&    769  \\
101&  12.05&    314&     0.29&   0.58&&   311&   0.08&    686  \\
102&  12.48&    628&     0.51&   0.56&&   598&   0.13&   1275  \\
103&  11.41&    505&     0.42&   0.51&&   515&   0.14&   1208  \\
107&  12.37&    684&     0.89&   0.42&&   666&   0.28&   1156  \\
111&  13.10&    487&     0.64&   0.61&&   478&   0.20&    736  \\
113&  12.32&    671&     0.74&   0.32&&   675&   0.44&    415  \\
121&  13.06&    599&     0.77&   0.33&&   606&   0.33&   1107  \\
124&  12.05&    766&     0.70&   0.19&&   751&   0.42&   1011  \\
132&  14.17&    447&     3.33&   0.43&&   403&   0.14&    316  \\
134&  12.68&    531&     0.66&   0.53&&   537&   0.25&   1017  \\
136&  13.29&    399&     0.81&   0.54&&   401&   0.14&    332  \\
146&  11.65&    689&     0.37&   0.32&&   735&   0.11&    759  \\
150&  11.21&    675&     0.09&   0.48&&   689&   0.04&    407  \\
157&  11.38&    389&     0.14&   0.53&&   393&   0.04&    600  \\
162&  11.63&    730&     0.26&   0.42&&   699&   0.09&    672  \\
175&  13.75&    630&     0.75&   0.50&&   636&   0.22&    437  \\
177&  13.30&    544&     1.17&   0.54&&   549&   0.26&    509  \\
180&  13.02&    555&     0.80&   0.63&&   543&   0.21&    567  \\
182&  12.71&    586&     0.34&   0.68&&   584&   0.13&    566  \\
186&  13.10&    275&     0.30&   0.63&&   280&   0.07&    574  \\
\enddata

\end{deluxetable}
\clearpage

%%%%%%%%%%%%%%%%%%%%%%%%%%%%%%%%%%%%%%%%%%%%%%%%%%%%%%%%%%%%%%%%%%%%%%%%%%%%%%%%%%

\begin{deluxetable}{cccc}
%\rotate
\tabletypesize{\scriptsize}
\tablecaption{
    Results of the fitted linear P-L relations for the 47 Semi-regular RSGs and LSP RSGs with
     distinguishable short period
    \label{pltab}
}
\tablewidth{0pt}
\tablehead{
    \colhead{Band} & \colhead{Slope $a$} & \colhead{intercept $b$} & \colhead{$\chi^{2}$}
}
\startdata
$V$      & -1.97$\pm$1.03& 18.13$\pm$2.85& 0.81\\
$J$      & -3.61$\pm$0.38& 18.91$\pm$1.06& 0.32\\
$H$      & -3.58$\pm$0.34& 18.02$\pm$0.95& 0.29\\
$K_{\rm S}$& -3.75$\pm$0.32& 18.13$\pm$0.88& 0.26\\
$[3.6]$& -3.98$\pm$0.29& 18.40$\pm$0.80& 0.24\\
$[4.5]$& -4.35$\pm$0.25& 19.41$\pm$0.68& 0.20\\
$[5.8]$& -4.54$\pm$0.28& 19.65$\pm$0.79& 0.23\\
$[8.0]$& -5.34$\pm$0.39& 21.43$\pm$1.10& 0.32\\
$[24]$ & -7.83$\pm$0.75& 26.34$\pm$2.14& 0.59\\
\enddata

\end{deluxetable}

%%%%%%%%%%%%%%%%%%%%%%%%%%%%%%%%%%%%%%%%%%%%%%%%%%%%%%%%%%%%%%%%%%%%%%%%%%%%%%%%%%

\begin{deluxetable}{ccc}
%\rotate
\tabletypesize{\scriptsize}
\tablecaption{
    The period and amplitude of the 51 LSP RSGs
    \label{lsptab}
}
\tablewidth{0pt}
\tablehead{
    \colhead{No.} & \colhead{Period (d)} & \colhead{Amplitude (mag)}
}
\startdata
3  & 2087&   0.23\\
7  & 2164&   0.07\\
20 & 2710&   0.08\\
22 & 2304&   0.03\\
25 & 2457&   0.20\\
27 & 2544&   0.11\\
33 & 1897&   0.05\\
35 & 1661&   0.34\\
46 & 2252&   0.08\\
48 & 2364&   0.15\\
53 & 1175&   0.06\\
64 & 2976&   0.03\\
69 & 2092&   0.10\\
73 & 2364&   0.09\\
74 & 2444&   0.17\\
75 & 1893&   0.07\\
76 & 2481&   0.23\\
80 & 1597&   0.08\\
84 & 2421&   0.12\\
87 & 2680&   0.14\\
93 & 1715&   0.08\\
96 & 2570&   0.07\\
100& 1587&   0.06\\
102& 2298&   0.23\\
104& 1342&   0.05\\
105& 2557&   0.07\\
106& 1579&   0.09\\
108& 2762&   0.05\\
114& 2958&   0.16\\
115& 1663&   0.06\\
120& 2433&   0.16\\
122& 2155&   0.16\\
125& 1248&   0.06\\
126& 1841&   0.10\\
127& 1811&   0.06\\
128& 1470&   0.13\\
129& 2314&   0.08\\
131& 2873&   0.09\\
133& 2816&   0.17\\
135& 2915&   0.09\\
147& 2040&   0.08\\
148& 2481&   0.13\\
152& 2976&   0.05\\
160& 2252&   0.17\\
161& 2832&   0.15\\
166& 902&    0.09\\
172& 2754&   0.23\\
177& 2232&   0.28\\
183& 1782&   0.04\\
184& 2624&   0.13\\
189& 1879&   0.12\\
\enddata

\end{deluxetable}
\clearpage

%%%%%%%%%%%%%%%%%%%%%%%%%%%%%%%%%%%%%%%%%%%%%%%%%%%%%%%%%%%%%%%%%%%%%%%%%%%%%%%%%%

\clearpage

%%%%%%%%%%%%%%%%%%%%%%%%%%%%%%%%%%%%%%%%%%%%%%%%%%%%%%%%%%%%%%%%%%%%%%%%%%%%%%%%%%

\clearpage

%%%%%%%%%%%%%%%%%%%%%%%%%%%%%%%%%%%%%%%%%%%%%%%%%%%%%%%%%%%%%%%%%%%%%%%%%%%%%%%%%%

\clearpage

%%%%%%%%%%%%%%%%%%%%%%%%%%%%%%%%%%%%%%%%%%%%%%%%%%%%%%%%%%%%%%%%%%%%%%%%%%%%%%%%%%

\clearpage

%%%%%%%%%%%%%%%%%%%%%%%%%%%%%%%%%%%%%%%%%%%%%%%%%%%%%%%%%%%%%%%%%%%%%%%%%%%%%%%%%%

\clearpage

%%%%%%%%%%%%%%%%%%%%%%%%%%%%%%%%%%%%%%%%%%%%%%%%%%%%%%%%%%%%%%%%%%%%%%%%%%%%%%%%%%

\clearpage

%%%%%%%%%%%%%%%%%%%%%%%%%%%%%%%%%%%%%%%%%%%%%%%%%%%%%%%%%%%%%%%%%%%%%%%%%%%%%%%%%%

\clearpage

%%%%%%%%%%%%%%%%%%%%%%%%%%%%%%%%%%%%%%%%%%%%%%%%%%%%%%%%%%%%%%%%%%%%%%%%%%%%%%%%%%

\clearpage

%%%%%%%%%%%%%%%%%%%%%%%%%%%%%%%%%%%%%%%%%%%%%%%%%%%%%%%%%%%%%%%%%%%%%%%%%%%%%%%%%%

\clearpage

%%%%%%%%%%%%%%%%%%%%%%%%%%%%%%%%%%%%%%%%%%%%%%%%%%%%%%%%%%%%%%%%%%%%%%%%%%%%%%%%%%

\clearpage

%%%%%%%%%%%%%%%%%%%%%%%%%%%%%%%%%%%%%%%%%%%%%%%%%%%%%%%%%%%%%%%%%%%%%%%%%%%%%%%%%%

\clearpage

%%%%%%%%%%%%%%%%%%%%%%%%%%%%%%%%%%%%%%%%%%%%%%%%%%%%%%%%%%%%%%%%%%%%%%%%%%%%%%%%%%

\clearpage

%%%%%%%%%%%%%%%%%%%%%%%%%%%%%%%%%%%%%%%%%%%%%%%%%%%%%%%%%%%%%%%%%%%%%%%%%%%%%%%%%%

\clearpage

%%%%%%%%%%%%%%%%%%%%%%%%%%%%%%%%%%%%%%%%%%%%%%%%%%%%%%%%%%%%%%%%%%%%%%%%%%%%%%%%%%

\clearpage

%%%%%%%%%%%%%%%%%%%%%%%%%%%%%%%%%%%%%%%%%%%%%%%%%%%%%%%%%%%%%%%%%%%%%%%%%%%%%%%%%%

\end{document}